\documentclass[useAMS,usenatbib]{mn2e}
\pdfoutput=1
\usepackage{lmodern}
\usepackage[utf8]{inputenc}
\usepackage[english]{babel}
\usepackage[T1]{fontenc}

\usepackage{rhmath}
\usepackage[nofloatbarriers]{rhgraphics}

\usepackage{mathrsfs}
\usepackage[flushleft]{threeparttable}

\usepackage[colorlinks=true,citecolor=blue,urlcolor=green]{hyperref}

\renewcommand{\eqref}[1]{Eq.~\ref{#1}}
\newcommand{\fref}[1]{Fig.~\ref{#1}}
\newcommand{\sref}[1]{\S\nobreak\ref{#1}}
\newcommand{\tref}[1]{Table~\ref{#1}}

\newcommand{\uHz}{\ensuremath{\umu\mathrm{Hz}}}
\newcommand{\numaxsun}{\ensuremath{\nu_{\mathrm{max},\sun}}\xspace}
\renewcommand{\sun}{\odot}
\newcommand{\Teffsun}{{T_{\mathrm{eff},\sun}}}
\newcommand{\Dpi}{{\Delta\Pi_1}}
\renewcommand{\epsilon}{\varepsilon}

\newcommand{\object}[1]{#1}

\graphicspath{{.}{./images/}}

\title[Peakbagging NGC~6819]{NGC~6819: testing the asteroseismic mass scale, mass loss, and evidence for products of non-standard evolution}
\author[R.~Handberg \etal]{R.~Handberg$^{1,2}$\thanks{E-mail: rasmush@phys.au.dk}, K.~Brogaard$^{2}$, A.~Miglio$^{1,2}$, D.~Bossini$^{1,2}$, Y.~Elsworth$^{1,2}$\newauthor D.~Slumstrup$^{2}$, G.~R.~Davies$^{1,2}$, W.~J.~Chaplin$^{1,2}$ \\
$^1$ School of Physics and Astronomy, University of Birmingham, Edgbaston, Birmingham B15 2TT, United Kingdom\\
$^2$ Stellar Astrophysics Centre (SAC), Department of Physics and Astronomy, Aarhus University, DK-8000 Aarhus C, Denmark}
\date{Received <date> / Accepted <date>}

\begin{document}
\maketitle

\begin{abstract}
We present an extensive peakbagging effort on \Kepler data of $\sim$50 red giant stars in the open star cluster NGC~6819. By employing sophisticated pre-processing of the time series and Markov Chain Monte Carlo techniques we extracted individual frequencies, heights and linewidths for hundreds of oscillation modes.

We show that the `average' asteroseismic parameter $\dnu{02}$, derived from these, can be used to distinguish the stellar evolutionary state between the red giant branch (RGB) stars and red clump (RC) stars.

Masses and radii are estimated using asteroseismic scaling relations, both empirically corrected to obtain self-consistency as well as agreement with independent measures of distance, and using updated theoretical corrections. Remarkable agreement is found, allowing the evolutionary state of the giants to be determined exclusively from the empirical correction to the scaling relations.
We find a mean mass of the RGB stars and RC stars in NGC~6819 to be $1.61\pm0.02\,\Msun$ and $1.64\pm0.02\,\Msun$, respectively. The difference $\Delta M=-0.03\pm0.01\,\Msun$ is almost insensitive to systematics, suggesting very little RGB mass loss, if any.

Stars that are outliers relative to the ensemble reveal overmassive members that likely evolved via mass-transfer in a blue straggler phase. We suggest that KIC~4937011, a low-mass Li-rich giant, is a cluster member in the RC phase that experienced very high mass-loss during its evolution. Such over- and undermassive stars need to be considered when studying field giants, since the true age of such stars cannot be known and there is currently no way to distinguish them from normal stars. 
\end{abstract}



\section{Introduction}\label{sec:intro}
Red giant stars in open clusters observed by the \Kepler mission \citep{Borucki2010,Borucki2016} have already been the subject of several asteroseismic studies \citep[see \eg][]{Basu2011,Stello2011,Corsaro2012,Miglio2012}. Asteroseismology of ensembles of giants with similar parameters and a common age and metallicity has allowed new insights into the clusters and stellar evolution, examples being asteroseismic membership information \citep{Stello2011}, mass loss on the red giant branch (RGB) \citep{Miglio2012} and stars with a non-standard evolutionary history \citep{Corsaro2012, Brogaard2015, Brogaard2016}. The asteroseismic studies of \Kepler cluster giants to date are based on so called average asterseismic parameters and in the most detailed case on measurements from collapsed echelle diagrams \citep{Corsaro2012} or by fitting an asymptotic expression \citep{Vrard2016}.
The next logical step, detailed peakbagging, meaning the extraction of individual oscillation modes and all their characteristics, has only been performed for a handful of stars in general and until now not for cluster members. The discipline has proven itself extensively for main-sequence and sub-giant stars \citep[see \eg][]{APTMCMC,PE15,Appourchaux2012,Appourchaux2014,Davies2016}, but has not been applied to a large extent to evolved red giants \citep{Corsaro2015}. This has mainly been due to complications introduced by the many mixed dipole modes. Here we present the first peakbagging effort on \Kepler light curves of evolved red giant stars in the open star cluster NGC~6819.

In \sref{sec:dataprep} and \sref{sec:global} we present the procedures used to prepare the \Kepler data for asteroseismic analysis and the methods used to extract global asteroseismic parameters from these. In \sref{sec:classification} we use asteroseismic information to distinguish between evolutionary states. In \sref{sec:peakbag} we go into detail on the peakbagging of individual oscillation modes, and in \sref{sec:averageseis} we use these results to derive new robust global oscillation parameters, and perform the same analysis for the Sun in \sref{sec:solar_reference}.
In \sref{sec:stellar_properties} and \sref{sec:nonstandard} we derive stellar properties and look into special cases of stars showing signs of non-standard evolution.
Finally we summarize the results and draw conclusions in \sref{sec:conclusion}.

\section{Data Preparation}\label{sec:dataprep}
The targets of our study are mainly those of \citet{Stello2011}, most of which were also studied by \citet{Corsaro2012}. We excluded a few stars for different reasons, being colours that were very significantly off compared to the cluster sequence in the colour-magnitude diagram (CMD), a high peak not related to solar-like oscillations in the power spectrum, or a very low \numax where measurements are extremely challenging ($\lesssim4\,\uHz$). The stars for which we made measurements are listed in \tref{tab:table1} and shown in the CMD of \fref{fig:hrd} based on photometry from \citet{Milliman2014,Hole2009} and \citet{Cutri2003} with labels as determined throughout this paper.  

\Kepler observations are divided into `quarters' of approximately 3 months duration due to the roll of the spacecraft required to keep the solar panels pointing towards the Sun. In this work we have used \Kepler data taken in long cadence (\SI{29.4}{min}) from  observing quarters Q0--Q16.
NGC~6819 falls on \Kepler CCD module 3, which failed after the first year of operation (Q4). This means that the time series of NGC~6819 is unfortunately missing data for 3 months every year, but have otherwise been near continuously observed.

The raw \Kepler data was preprocessed using the procedures described in \citet{KASOCFilter}.
Normally this procedure involves using the \Kepler target pixel data to automatically redefine the pixel masks used for extracting the stellar flux using simple aperture photometry. This is to allow more flux from the target through the aperture and has been shown for field stars to improve asteroseismic analyses compared to the \Kepler standard masks. However, since we are here dealing with a relatively crowded field, this procedure was not ideal for all targets and all timeseries were therefore visually inspected to decide if the new or original masks performed better for each target with respect to contamination and overall scatter.

From initial analyses of the data and previously published asteroseismic parameter values \citep{Stello2011}, we had initial indications of the frequency ranges in which the stars oscillate, and care was taken to scale the filter parameters in such a way that the oscillation signals would not be perturbed in any way \citep[see][]{KASOCFilter}.

\section{Global asteroseismic parameters}\label{sec:global}
The acoustic oscillation frequencies of solar-like stars are approximatively described by \citep{Tassoul1980}:
	\begin{equation}\label{eqn:asymptotic}
		\nu_{n\ell} \simeq \Dnu (n + \ell/2 + \epsilon) - \dnu{0\ell} \, ,
	\end{equation}
where $n$ and $\ell$ denote the radial order and spherical degree of the oscillation mode respectively, $\Dnu$ is the large frequency separation, $\epsilon$ is an offset introduced by surface effects and $\dnu{0\ell}$ is the small frequency separation.

The power spectrum of a solar-like oscillator is dominated by an overall background arising primarily from the turbulent convective motions in the outer layers of the star. This convective motion drives the acoustic oscillations in the star, often referred to as the $p$-modes of oscillation. The measured power spectrum of a solar-like oscillator will therefore consist of individual peaks from each mode of oscillation sitting on top of the granulation background signal. These modes will exhibit a regular pattern with the distance between consecutive overtones of approximately $\Dnu$.
The first step in the analysis of a solar-like oscillator is to fit and describe the background signal. However, since there does not exist any physically founded description for the true shape and size of the granulation signal, it is common practice to turn to an empirical description.
Historically, following the principles in \citet{Harvey1985}, the background is described as the sum of several Lorentzian (or super- or pseudo-Lorentzian) functions. This choice stems from the empirical description of the brightness of the granules that falls of exponentially.

In this work we have opted for not selecting \emph{a priori} any particular model for the background, but instead the backgrounds are fitted, using maximum a posteriori probability estimation, using four different and commonly used descriptions of the background signal \citep{Harvey1985,KaroffThesis,Kallinger2012,Karoff2013,Kallinger2014}:
\begin{align}
	N(\nu) &= \eta(\nu) \cdot \sum_{k=1}^2 \frac{\xi\sigma_k^2\tau_k}{1 + (2\pi\nu\tau_k)^2} + K \label{eqn:mdl:2}\\
	N(\nu) &= \eta(\nu) \cdot \sum_{k=1}^2 \frac{\xi\sigma_k^2\tau_k}{1 + (2\pi\nu\tau_k)^4} + K \label{eqn:mdl:4}\\
	N(\nu) &= \eta(\nu) \cdot \sum_{k=1}^2 \frac{\xi\sigma_k^2\tau_k}{1 + (2\pi\nu\tau_k)^2 + (2\pi\nu\tau_k)^4} + K \label{eqn:mdl:24}\\
	N(\nu) &= \eta(\nu) \cdot \sum_{k=1}^2 \frac{\xi\sigma_k^2\tau_k}{(1 + (2\pi\nu\tau_k)^2)^2} + K \label{eqn:mdl:22}
\end{align}
where $\sigma_k$ and $\tau_k$ is the amplitudes and time-scales of each component, $\xi$ is a normalisation constant, and $K$ is the white noise level. The frequency dependent factor $\eta(\nu)\equiv\sinc^2(\Delta T_\mathrm{int}\cdot\nu)$ is the attenuation arising from gathering the data over a time-span, $\Delta T_\mathrm{int}$ (the exposure time), which in the case of \Kepler long cadence data is \SI{1765.5}{s}. $\sinc$ is the normalized sinc-function: $\sinc(x)=\sin(\pi x)/(\pi x)$.
In addition to this comes a Gaussian envelope accounting for the oscillation power. The location of the highest point in this envelope is defined as $\numax$, and the resulting combined model power spectrum is defined as follows:
	\begin{equation}\label{eqn:limitspectrum}
		\mathscr{P}(\nu) = N(\nu) + \eta(\nu) \, a_\mathrm{env}\exp\kant*{\frac{-(\nu - \numax)^2}{2\sigma_\mathrm{env}^2}} \, ,
	\end{equation}
where $a_\mathrm{env}$ and $\sigma_\mathrm{env}$ is the height and width of the envelope respectively.
This model is fitted to the full power spectrum. The large separation, $\Dnu_\mathrm{global}$, is then estimated by the use of the power spectrum of the power spectrum \citep[PSPS; see][]{HandbergThesis} in the frequency range $\numax \pm 2\sigma_\mathrm{env}$. The model is fitted with constraints based on initial guesses on \numax and \Dnu (from visual inspection) and in each case, the background fit and large separation calculation are iterated three times to yield a consistent result. Finally the background with the lowest Bayesian Information Criterion \citep[BIC; ][]{Schwarz1978} is chosen as the adopted background:
	\begin{equation}
		\mathrm{BIC} = -2 \ln\mathcal{L} + k\cdot\ln N \, ,
	\end{equation}
where $\mathcal{L}$ is the calculated maximum likelihood from the best fit, $k$ is the number of degrees of freedom and $N$ is the number of points in the power spectrum being fitted. It should be noted that the BIC is an approximation to the true Bayesian evidence, and only valid when the data distribution is in an exponential family, but this is also the case for power spectra.
Once the optimal shape of the background has been chosen, the fitted parameters are passed as first guesses to a Markov chain Monte Carlo method \citep{APTMCMC} which yields the final values of the parameters and corresponding error estimates.

The reason for not choosing a fixed background model \emph{a priori} is that, prior to comparing with the data, we do not have any preference for any specific empirical model. Fixing the background to a single model for all stars comes with the prior assumption that the exact shape of the granulation background is well known and static for all stars -- a statement which appears premature, although our understanding of granulation has significantly progressed in recent years \citep[see \eg][]{Trampedach2013,Trampedach2014}.

\begin{figure}
	\centering
	\includegraphics[width=\columnwidth]{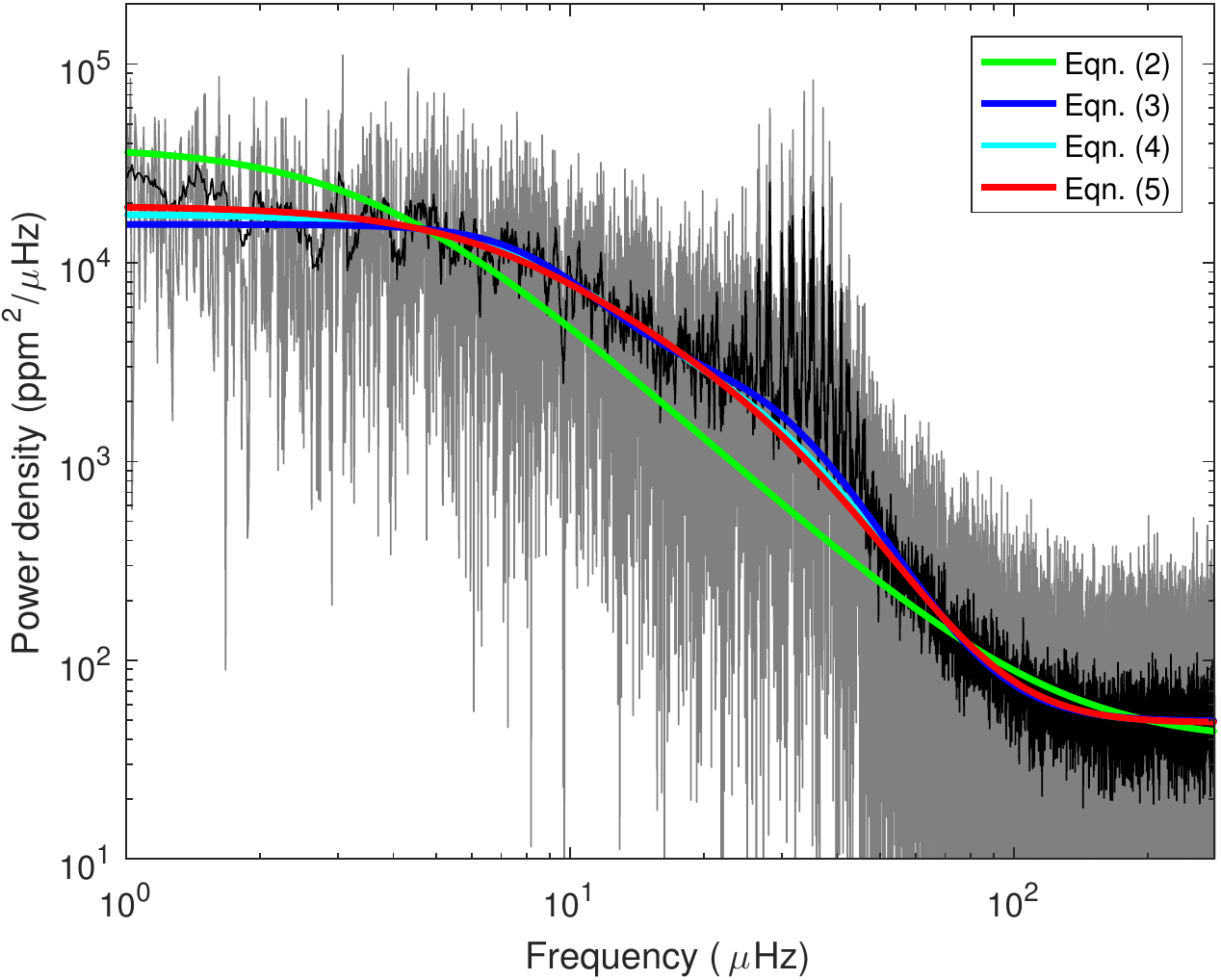}
	\caption{Power density spectrum and all the fitted background models for KIC~4937576 ($\numax\!=\!33\,\uHz$). In gray the original power density spectrum and in black the same spectrum smoothed with a \SI{0.14}{\uHz} wide boxcar. \eqref{eqn:mdl:24} was the preferred background model for this star.}
	\label{fig:many-backgrounds}
\end{figure}

\fref{fig:many-backgrounds} shows the power density spectrum of KIC~4937576 with all the different background models considered overplotted. The preferred background model for this particular target is \eqref{eqn:mdl:24}. What can also be seen from this is that Eqns.~\ref{eqn:mdl:4}--\ref{eqn:mdl:22} are very similar, but \eqref{eqn:mdl:2} reproduces the background-level very poorly in this case.

\begin{figure}
	\subfloat[\label{fig:best_backgrounds}]{\includegraphics[width=\columnwidth]{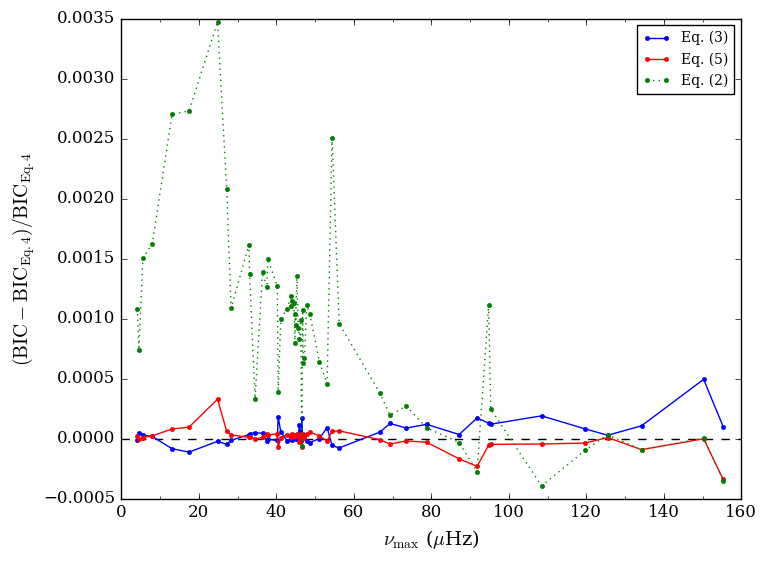}}\\%
	\subfloat[\label{fig:background_comparison}]{\includegraphics[width=\columnwidth]{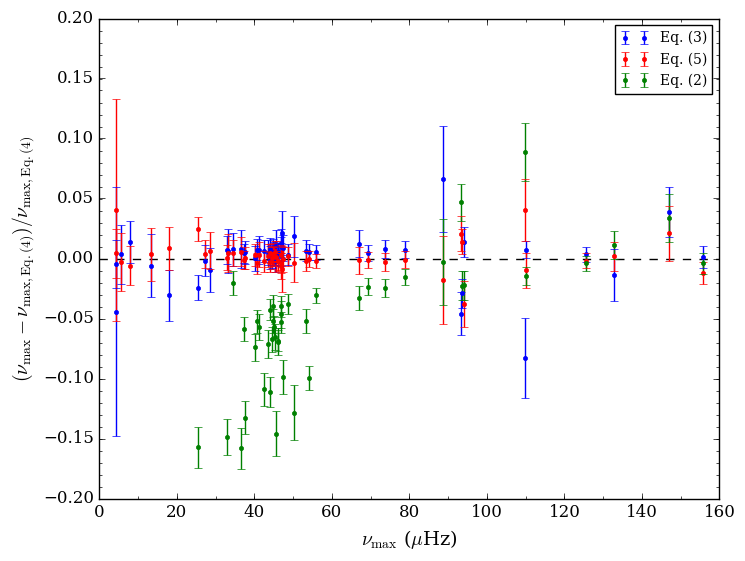}}%
	\caption{\protect\subref{fig:best_backgrounds}: Bayesian Information Criterion (BIC) for different background models. For each star, the background model with the lowest BIC was used in the following analysis. \protect\subref{fig:background_comparison}: Comparison of $\numax$ for different background models.}
\end{figure}

\fref{fig:best_backgrounds} shows the BIC for each background model for all stars sorted by $\numax$. The model with the lowest value of BIC was used in the further analysis. As seen from \fref{fig:best_backgrounds}, there seems to be a slight correlation between which background model is preferred as a function of stellar evolution ($\numax$). Disregarding \eqref{eqn:mdl:2}, which overall does a poor job of describing the power spectra, \eqref{eqn:mdl:4} seems to perform better at low \numax until roughly the location of the red clump (RC) ($\numax$=30--50\,\uHz) and at higher \numax \eqref{eqn:mdl:22} is in general preferred. \eqref{eqn:mdl:24} in general lies between these two in terms of BIC.
Of course, care has to be taken in drawing too firm conclusions from this, since the BIC is only a measure of how well the models describe a given dataset. It does not necessarily say anything about the validity of the background models in general.

Since the most important parameter of these fits is $\numax$, a very important question then arises: Does the determined value of $\numax$ depend on which background model we use?
We have therefore fitted the power spectra using the different background-prescriptions, but without selecting only the best one. In \fref{fig:background_comparison} the relative difference between the returned $\numax$ are shown. It is clear that there are significant deviations between \eqref{eqn:mdl:2} and the other models at low $\numax$. But since this model is never selected for low $\numax$ this has no impact on the returned $\numax$. In general the differences between the returned values of $\numax$ are almost all consistent with zero within the errorbars. This means that the values of $\numax$ are largely independent of which background model is used, with the exception of \eqref{eqn:mdl:2}.

We also tested the background model proposed by \citet{Kallinger2014}. This model is identical to \eqref{eqn:mdl:4}, but with the addition of an ``instrumental'' component, increasing the number of free parameters by two, which is penalised in the BIC. In the best cases it performed as well as \eqref{eqn:mdl:4}, but often significantly worse, particularly at low \numax. We also observed larger deviations in \numax when using this model when compared to the other models. For a subset of we also observed problems of getting the fit to converge. We also ascribe this to the additional free parameters.

\section{Period spacings and classifications}\label{sec:classification}
The evolutionary state classification is mainly based on the appearance of the mixed-mode structure of the dipole modes in the power spectrum. The procedure is described in detail in \citet{Elsworth2016}, but the outline is as follows.
 
We use the previously determined value of \Dnu to divide the spectrum into zones in which either odd or even $\ell$ modes are located. Within each zone we then use statistical tests on the unsmoothed spectrum to pick out significant spikes. The significance threshold is set so as to exclude almost all the background noise but to pick up small features in the spectrum.

This is a frequentist approach that makes no assumptions about what will be present in the data. The result of the test is a set of frequencies at which spikes have been found.
The criterion employed is that a feature is considered significant if there is a less than 20\% probability of a false detection over a frequency range of $3.5\uHz$. This range represents a typical \Dnu for a RC star.
 
In each zone, we find the separation in frequency of each feature from all other features. Thus, if there were $x$ features identified from the statistical tests there would be $x(x-1)$ frequency differences. A histogram is formed from these which gives the number of occurrences of different frequency differences. The features to be found in these histograms carry information about the features within the spectrum and are different in the two zones of odd and even $\ell$. The classification uses just the odd $\ell$ zone. The histogram of the odd-$\ell$ zone will show the large frequency separation but will additionally show features that are a consequence of the presence of mixed modes.

Core-helium burning stars tend to show several mixed modes of similar heights over many orders with observed period spacings that typically lie between about \SI{100}{s} and \SI{250}{s}. For RGB stars, the period spacing is much smaller at around \SI{60}{s} and the mixed modes have a sufficiently large inertia that they are not always visible \citep[see \eg][]{Bedding2011,Mosser2012}. These properties influence the features in the frequency difference histogram.
Furthermore, by definition, the asymptotic period spacing is uniform in period and not in frequency. This means that the frequency difference between two mixed modes at the low-frequency end of the spectrum is significantly smaller than the difference at the high-frequency end. This variation influences the width of the feature in the frequency-separations which, expressed as a function of \Dnu, can be used to differentiate the two classes of red giants. More details on the method are given in \citet{Elsworth2016}.

Stars which have depressed dipole modes cannot be classified in this way. Therefore, a few stars were instead classified by their CMD positions and according to their small spacing $\delta\nu_{02}$, see \ref{sec:averageseis}.

\begin{figure}
	\centering
	\includegraphics[width=\columnwidth]{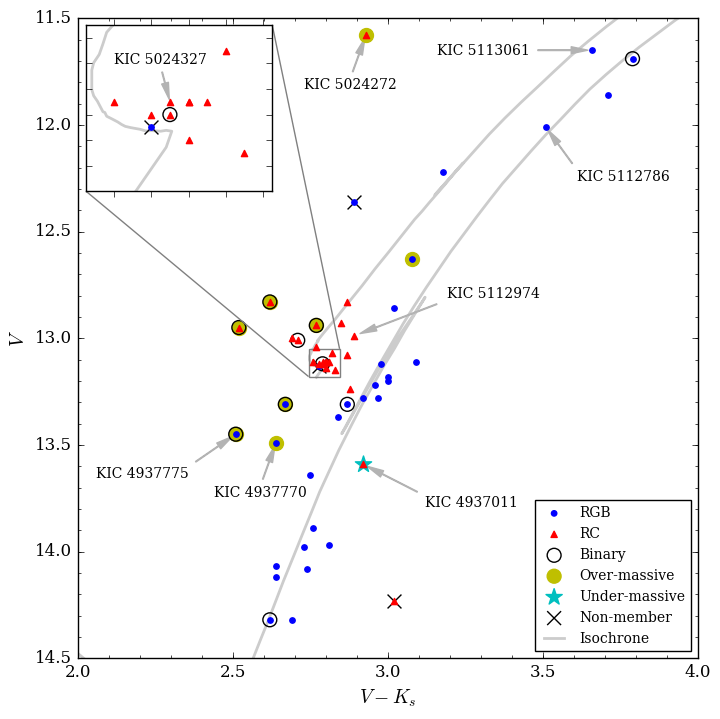}
	\caption{Colour-magnitude diagram of the observed giants in NGC~6819. The isochrone is from PARSEC \citep{Bressan2012}.}
	\label{fig:hrd}
\end{figure}

\section{Peakbagging}\label{sec:peakbag}
\begin{figure*}
	\centering
	\subfloat[KIC~5112072. Low Red Giant Branch star.]{\includegraphics[width=\textwidth]{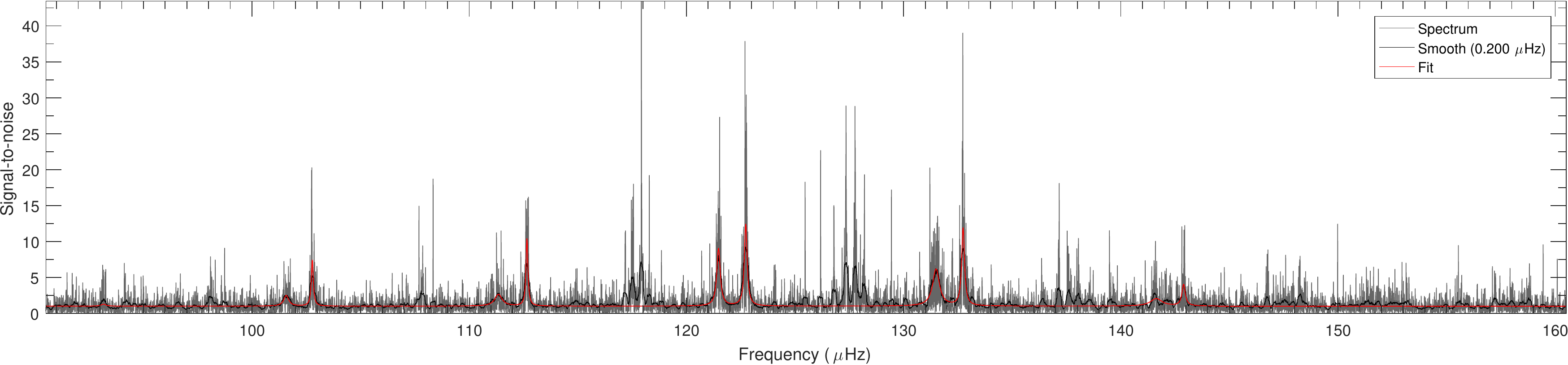}}\\%
	\subfloat[KIC~5024043. ``Clean dipole'' star.]{\includegraphics[width=\textwidth]{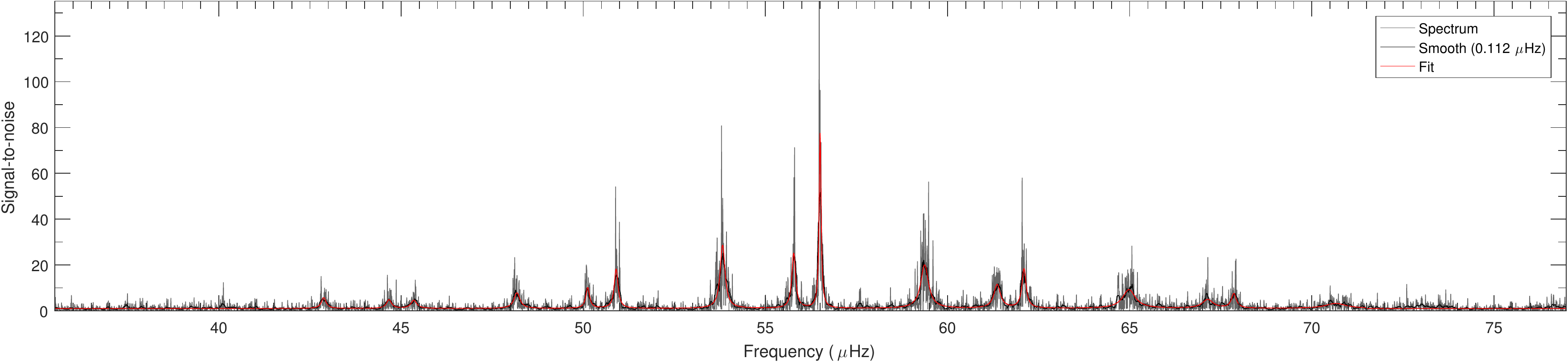}}\\%
	\subfloat[KIC~5024327. Red Clump star.]{\includegraphics[width=\textwidth]{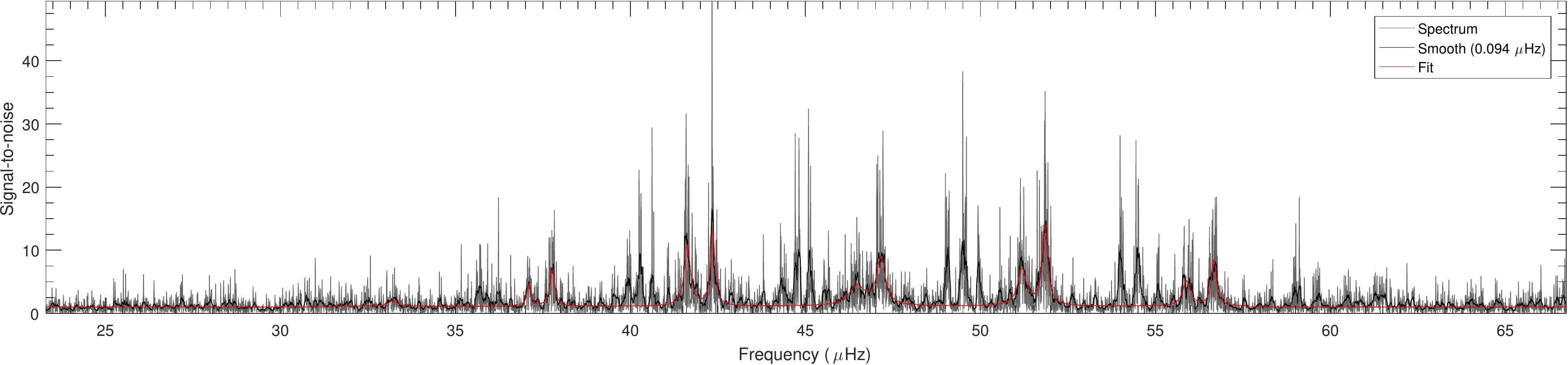}}
	\caption{Power spectra of representative stars of the three empirical types of stars in NGC~6819. Power spectra have here been divided by the fitted background model. The red line indicates the fitted limit spectrum (\eqref{eqn:limitspectrum}), using the medians of the posterior probability distribution.}
	\label{fig:groups}
\end{figure*}

From a simple look at the power spectra of the stars in NGC~6819 and in particular on the structure of the dipole modes of these, it becomes apparent that these can be empirically divided into three groups: Low RGB, RC, and ``clean dipole'' stars. In \fref{fig:groups} the power spectra of representative stars for each of the three groups are shown. The low RGB stars, which are stars with $\numax\gtrsim 60\uHz$, are characterized by having rich spectra with many mixed dipole modes. In some of them rotational splitting is visible, but reasonably distinguishable from the mixed mode structure as the rotational splittings are still smaller than the observed period spacings.
For the RC stars, which lie in the region $\numax\approx$35--45$\uHz$, the power spectra are also very rich in structure. However, for these stars the asymptotic dipole structure is not always as clear. This may simply be because rotational splitting, which is now on the same scale as the period spacings, is complicating the spectra.
The last group, the ``clean dipole'' stars, are evolved RGB stars with $\numax\lesssim 60\uHz$ and power spectra which are very reminiscent of main-sequence stars. Each order apparently only contains a single dipole mode with a shape similar to the radial modes, sitting roughly where you would expect it from the asymptotic description (\eqref{eqn:asymptotic}). From stellar models we know that these dipole modes are in fact composed of several mixed dipole modes sitting very close together in frequency around the \emph{nominal $p$-mode}, $\nu_p$, meaning the frequency where the pure $p$-mode would sit if there was no mixing with $g$-modes. Each individual mode is not resolved in the power spectrum, but the combined power excess from them yields what appears as a single clean Lorentzian mode \citep[see \eg][]{Dupret2009,Montalban2010}. 

For the different empirical groups of stars, we opted for slightly different strategies in which modes were included in the peakbagging. In all cases we included a single oscillation mode for each $\ell\!=\!0$ and $\ell\!=\!2$ modes. We know that in principle the $\ell\!=\!2$ may also contain mixed modes, but these are in general confined close together and hence unresolved. Both the mixed modes signatures and any rotational splitting will manifest in nearly symmetrical features in frequency and will not cause significant errors in the determined frequencies. It is however important to keep this in mind when comparing the parameters to stellar models and we add a note of caution that individual $\ell\!=\!2$ modes can be biased because of this.

Following the same line of thought, in the cases of the ``clean dipole'' stars, we have elected to fit the dipole modes with a single Lorentzian peak with frequency, linewidth and height as free parameters, but not including rotational splitting.

For the low RGB and RC stars, the regions between the $\ell=0,2$ mode pairs containing power from the dipole modes was taken out. This was done by simply not including these regions in the sum over all bins in the calculation of the log-likelihood between the model spectrum and the observed power spectrum.
It is clear that we could include the individual mixed dipole if a clear mode identification could be obtained. In general this is very difficult in the red clump stars, but can be done for some of the low RGB stars.
For cases where it was possible, for low RGB stars, we chose to include the individual dipole modes in the peakbagging, using the asymptotic relation from \cite{Mosser2012} for the frequencies of mixed dipole modes, $\nu_\mathrm{mix}$, to set initial guesses. The relation which needs to be iteratively solved for $\nu_\mathrm{mix}$ is the following:
	\begin{equation}\label{eqn:mixed_modes}
		\nu_\mathrm{mix} \simeq \nu_p + \frac{\Dnu}{\pi} \arctan\kant*{ q \tan\p*{\frac{1}{\Dpi\nu_\mathrm{mix}}} - \epsilon_g } \, ,
	\end{equation}
where $\nu_p$ is the nominal $p$-mode frequency, $q$ is the coupling, $\Dpi$ is the asymptotic dipole period-spacing and $\epsilon_g$ is a phase shift.
In addition to this, it is essential to include rotational splitting in order to describe the oscillation spectrum. We have included this following \cite{Goupil2013}, where the rotational splitting of a mixed mode at $\nu_\mathrm{mix}$ is given by:
	\begin{equation}\label{eqn:mixed_modes_rotation}
		\delta\nu_s \approx \p*{ \frac{1-2\mathcal{R}}{1+4\frac{\Dpi\nu_\mathrm{mix}^2}{\Dnu} \cos^2\p*{\frac{\pi}{\Dpi\nu_\mathrm{mix}}}} + 2\mathcal{R} } \delta\nu_{s,\mathrm{max}} \, ,
	\end{equation}
where $\mathcal{R}$ is defined as the ratio between the average rotation rate of the envelope and the core, $\mathcal{R} \equiv \mean{\Omega}_\mathrm{env}/\mean{\Omega}_\mathrm{core}$, and $\delta\nu_{s,\mathrm{max}}$ is the maximal rotational splitting of the most $g$-like modes.
In order to find the values of the free parameters in \eqref{eqn:mixed_modes} and \eqref{eqn:mixed_modes_rotation} which best reproduce the observed spectrum, parameters were varied manually until good agreement was found between the peaks of the observed spectrum and the calculated frequencies. Once initial guesses were set using the above prescription, frequencies, linewidths and rotational splittings were fitted individually for all modes. See also \citet{Corsaro2015} who performed peakbagging of this type of stars.

We fit the individual oscillation modes using the Markov chain Monte Carlo (MCMC) techniques described in \citet{APTMCMC}.
The principle is to fit the observed power density spectrum with a model power spectrum that is defined as a simple sum of Lorentzian profiles and the background contribution:
\begin{equation}
	\mathscr{P}(\nu) = \eta(\nu) \cdot \sum_{n,\ell}\sum_{m=-\ell}^\ell \frac{\mathcal{E}_{\ell m}(i) H_{n\ell}}{1 + \tfrac{4}{\Gamma_{n\ell}^2}(\nu - \nu_{n\ell} - m\delta\nu_{s,n\ell})^2} + N(\nu) \, ,
\end{equation}
where mode frequencies, heights, linewidths and rotational splittings ($\nu_{n\ell}$, $H_{n\ell}$, $\Gamma_{n\ell}$, $\delta\nu_s$) are all treated as independent free parameters. In the cases where rotation was not included in the fit, $\delta\nu_s\!=\!0$ and $\mathcal{E}_{\ell m}(i)\!=\!1$ were simply fixed.

Priors on mode frequencies and linewidths are set as uniform distributions (boundaries of which are defined on per-case basis). The prior on heights are set as modified Jeffreys priors \citep[see \eg][]{APTMCMC}. The parameters for the background are also treated as free parameters, but with strong priors set as the posterior probability distributions of the background fit already made to the full frequency range. 

After the fitting is done, all posterior probability distributions are inspected manually to reject any modes that could not be reliably fitted. For each posterior, the median value is chosen as the central value and the 1$\sigma$ errorbars are calculated from the 68\% credible region. This means that there is 68\% confidence that the parameter will have its true value in this interval.

We did not correct the final oscillation frequencies for the line-of-sight Doppler velocity shifts as described in \citet{Davies2014}. Although the radial velocity measurements are available for the individual stars, we are dealing with low-frequency oscillations and the cluster system velocity of only 2.45$\pm$1.02\,km/s \citep{Milliman2014} is low, rendering the corrections insignificant.

\section{Extracted average seismic parameters}\label{sec:averageseis}
Our goal is now to use the individual frequencies we have extracted from the detailed peakbagging to reconstruct the average seismic parameters of \eqref{eqn:asymptotic}. This has the advantage that the same procedures can easily be applied to frequencies originating from stellar models, allowing for direct comparison. This is something that is, for example, not easily achievable with $\Dnu_\mathrm{global}$ derived from the autocorrelation of the measured power spectrum.
From the extracted $\ell=0,2$ frequencies we can construct the average parameters $\Dnu_\mathrm{fit}$, $\epsilon$ and $\delta\nu_{02}$.
The first two are found by a weighted linear fit to the extracted radial mode frequencies, where the weights are given as $\sigma_i^{-2}$, where $\sigma_i$ is the uncertainty of the individual frequencies. The slope of the fitted curve directly gives $\Dnu_\mathrm{fit}$, while the intersection with the zero-axis yields $\epsilon$.
$\mean{\dnu{02}}$ is found as the weighted mean of the frequency differences between adjacent $\ell=2$ and $\ell=0$ modes.

\begin{figure}
\centering
\includegraphics[width=\columnwidth]{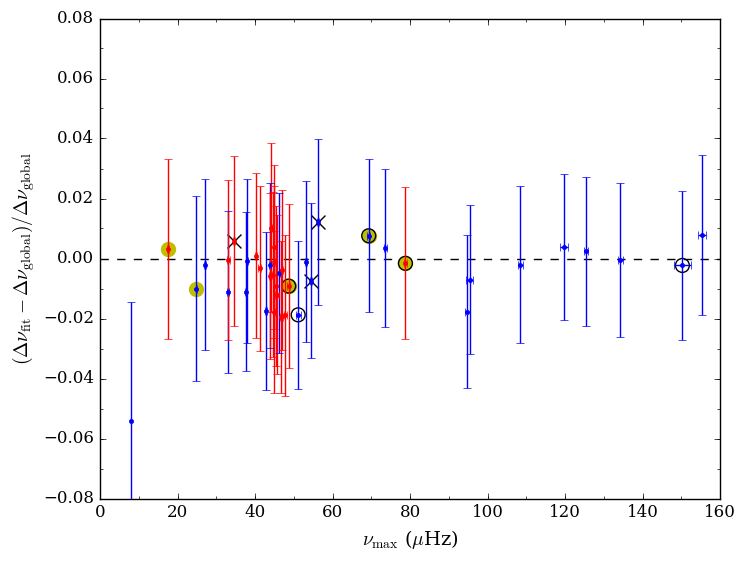}
\caption{Comparison between $\Dnu_\mathrm{fit}$ and $\Dnu_\mathrm{global}$ in the cases where they could both be obtained. Colours and points are the same as in \fref{fig:hrd}. This clearly shows that there is no significant difference between the two estimates, however some scatter on the $\sim$1\% level is seen.}
\label{fig:lsep_comparison}
\end{figure}

In \fref{fig:lsep_comparison} the values of $\Dnu_\mathrm{fit}$, determined from the measured individual frequencies, are compared to the values of $\Dnu_\mathrm{global}$, which as previously mentioned is measured using the PSPS. In general there is good agreement between the two estimates, which is reassuring for stars where individual frequencies can not be obtained. We do see some scatter at the 1\% level, but well within the errorbars of the measurements.

As expected we find a strong correlation between the observed $\Dnu$ and $\numax$ \citep[see \eg][]{StelloScalingRelation,Huber2011}. The scaling between the parameters obtained in this work, assuming the usual power-law dependence, is the following:
	\begin{equation}\label{eqn:dnu_numax_scaling}
		\Dnu_\mathrm{fit} = (0.248\pm0.009) \cdot \numax^{(0.766\pm0.008)} \, .
	\end{equation}
Comparing to others, it is important to note that this scaling is for a particular mass (the RGB mass of NGC~6819, which we derive later) and metallicity.

As previously mentioned, the second global asteroseismic parameter that we can estimate from the individual frequencies of the radial modes is the $\epsilon$ parameter, often refereed to as the ``surface-term''.
Following \citet{MosserUniversalPattern} we determine the best fitting power function of the form $\epsilon=a+b \log_{10}(\Dnu)$, which yields the following scaling relation for $\epsilon$:
\begin{equation}\label{eqn:dnu_epsilon_scaling}
\epsilon = (0.634{\pm}0.040) + (0.612{\pm}0.048) \cdot \log_{10}(\Dnu_\mathrm{fit}) \, .
\end{equation}
The parameters of this fit are within the 1-$\sigma$ errors of a similar scaling relation reported by \citet{Corsaro2012}.

\begin{figure}
\centering
\includegraphics[width=\columnwidth]{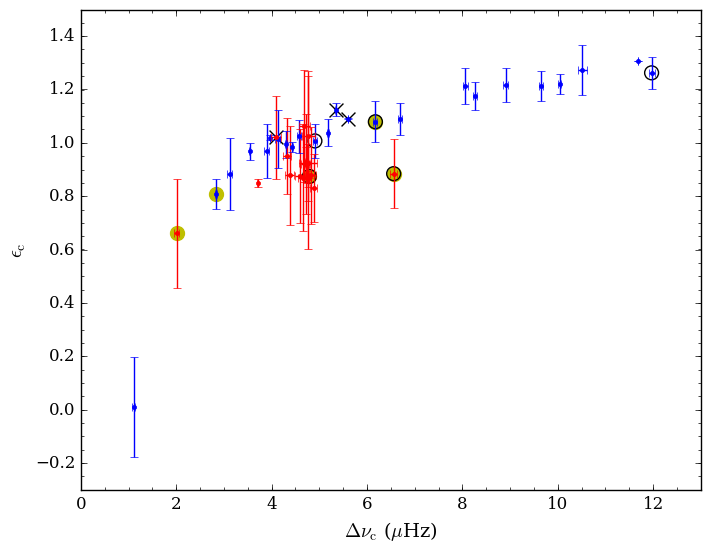}
\caption{Seismic $\epsilon_c$ vs $\Dnu_c$ parameters determined from the three central, individually fitted, radial modes. Colours and points are the same as in \fref{fig:hrd}. The reason for the, in general, larger errorbars on the RC stars is fact that RC stars have a more `messy' power spectrum, as seen in \fref{fig:groups}, which translates into larger errorbars on the radial mode frequencies.}
\label{fig:epsilon}
\end{figure}

\citet{Kallinger2012} proposed and alternative definition where, unlike in the above mentioned prescription, only the three central radial modes closest to $\numax$ are used to measure $\Dnu_c$ and $\epsilon_c$ (here indicated by the subscript). \citet{Kallinger2012} used this definition to determine the evolutionary phase of field stars. Both definitions of $\Dnu$ and $\Dnu_c$, and of $\epsilon$ and $\epsilon_c$ gives very similar results but, not surprisingly, the errorbars on $\epsilon$ are remarkably smaller, at the cost of not tracing the evolutionary differences. However, as we show in \fref{fig:epsilon}, $\epsilon_c$ is in our case also not able to separate the RGB stars from the RC stars particularly well. We do see the trend reported by \citet{Kallinger2012} that RC stars tend to have a smaller $\epsilon_c$ at $\Dnu\sim$4--4.5$\uHz$, but the distinction does not seem to be significant on an individual star basis due to a significant overlap in $\epsilon_c$ for the RC and RGB stars.

\subsection{Small frequency separation}
\begin{figure}
	\centering
	\subfloat[\label{fig:d02:sub:ratio}]{\includegraphics[width=\columnwidth]{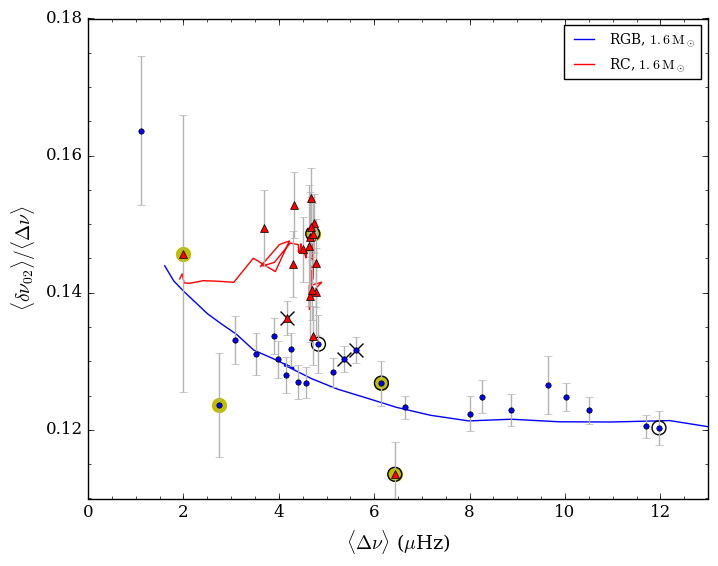}}\\\vspace*{-1.1em}%
	\subfloat[\label{fig:d02:sub:models}]{\includegraphics[width=\columnwidth]{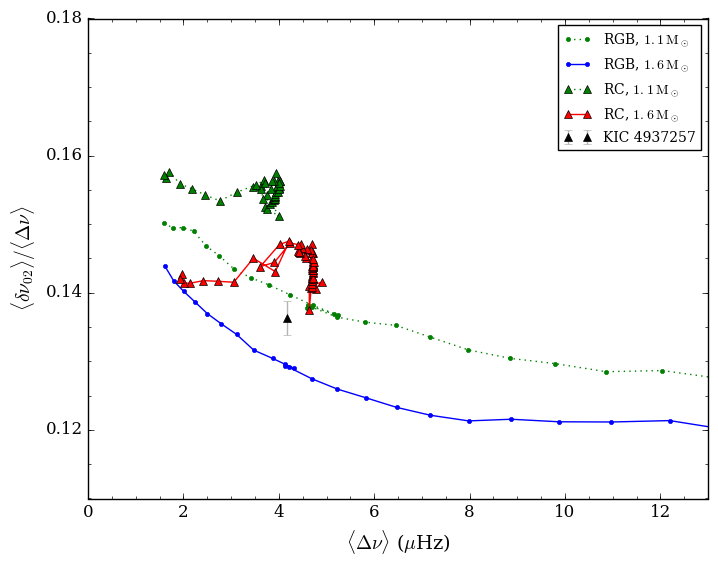}}\\\vspace*{-1.1em}%
	\subfloat[\label{fig:d02:sub:rgb}]{\includegraphics[width=\columnwidth]{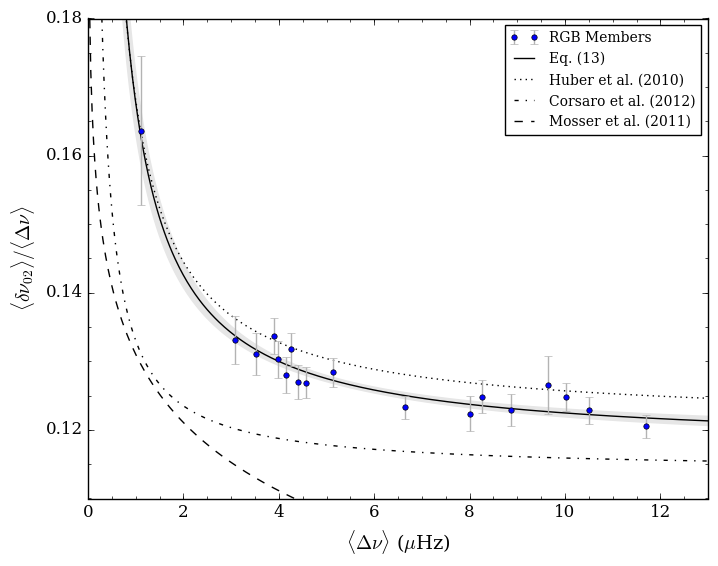}}%
	\caption{Small frequency separation as a function of large frequency separation. \protect\subref{fig:d02:sub:ratio}: Measured values for all stars overlaid with predictions from stellar models of $1.6\,\Msun$. \protect\subref{fig:d02:sub:models}: Predictions from stellar models for two different masses. \protect\subref{fig:d02:sub:rgb}: Comparison between empirical scaling relations and RGB single cluster members. The shaded gray region denotes the 1-$\sigma$ confidence interval around the fitted curve.}
	\label{fig:d02}
\end{figure}

However, if we turn our attention to the parameter $\mean{\dnu{02}}/\Dnu_\mathrm{fit}$ \citep[see \eg][]{Montalban2010a,Montalban2010,Montalban2011}, the distinction between RGB and RC stars turns out to be more prominent.
In \fref{fig:d02:sub:models} this quantity is shown as a function of the large frequency separation determined from stellar models and in \fref{fig:d02:sub:ratio} the same is shown for the measured values.
It would appear the we have identified an alternate way to distinguish the RGB from the RC phase, since our measurements clearly separate the two evolutionary phases. As seen, this is also in very good agreement with our theoretical predictions for the red giant mass of NGC~6819 (see~\sref{sec:stellar_properties}). However, while our model predictions for a lower mass of $1.1\,\Msun$ suggest a similar difference in $\mean{\dnu{02}}/\Dnu_\mathrm{fit}$ between the two evolutionary phases, the measurements by \citet{Corsaro2012} of giants in the old open cluster NGC~6791 showed an overlap where the $\mean{\dnu{02}}/\Dnu_\mathrm{fit}$ of RC stars scattered on both sides of the RGB stars (see panel b of their Fig. 4). Since NGC~6791 is very metal-rich one could suspect this to be the cause of the disagreement. Luckily, our analysis of individual targets in \sref{sec:nonmembers} identified a non-member RC star of solar metallicity with a mass of $1.1\,\Msun$ that allows us to test that hypothesis. We show that one measurement in \fref{fig:d02:sub:models} where it demonstrates the same issue; the value of $\mean{\dnu{02}}/\Dnu_\mathrm{fit}$ does not follow the model prediction for a $1.1\,\Msun$ RC star of solar metallicity. Although outside the scope of the present paper this clearly should be investigated further.

In \fref{fig:d02:sub:rgb} the solid line corresponds to a linear fit to the RGB stars given by the following expression:
\begin{equation}\label{eqn:dnu_d02_scaling}
\mean{\dnu{02}} = (0.0504{\pm}0.0074) + (0.1179{\pm}0.0014) \cdot \Dnu_\mathrm{fit} \, ,
\end{equation}
plotted together with previously published scaling relations by \citet{Huber2010}, \citet{MosserUniversalPattern} and \citet{Corsaro2012}. The difference in $\mean{\dnu{02}}$ between RGB and RC stars is similar to previous studies (see \fref{fig:d02}), but there are systematic offsets in the absolute numbers, which is most likely due to the different methods used to measure $\mean{\dnu{02}}$.

\begin{figure}
	\centering
	\includegraphics[width=\columnwidth]{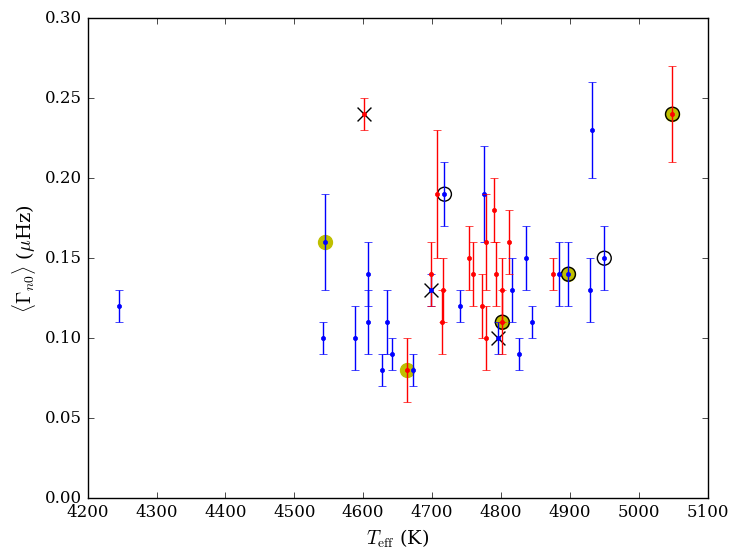}
	\caption{Average linewidth of radial modes as a function of effective temperature of the stars. Symbols have the same meaning as in previous figures. We do not see a significant correlation like reported by \citet{Corsaro2012}. The effective temperatures are derived in \sref{sec:stellar_properties}.}
	\label{fig:teff_linewidth}
\end{figure}

We have also looked at the extracted average linewidth of the radial modes, and come to the conclusion that this shows no clear systematic variations as a function of evolutionary state. All stars have $\mean{\Gamma_{n0}}$ in the range 0.08--0.2\,\uHz. There could maybe be a very slight variation of the linewidth as a function of effective temperature as reported by \citet{Corsaro2012}, but the correlation is not statistically significant from our given sample. See \fref{fig:teff_linewidth}. The effective temperatures are derived in \sref{sec:stellar_properties}.

We add a note of caution that the measured linewidths of course will depend on the chosen noise background formulation \citep[see \eg][]{Appourchaux2012,Appourchaux2014,Corsaro2014}.

\section{Solar reference values}\label{sec:solar_reference}
In order to utilize scaling relations for the stellar parameters (see next section), we need values for $\numax$ and $\Dnu$ of the Sun using similar methods to the ones used to obtain the values from the stars. To do this for $\numax$ we fitted a 12 year power spectrum from the green channel of the VIRGO instrument aboard the SoHO spacecraft with the prescription given in \sref{sec:global}. This yields a solar value for $\numaxsun = \SI{3090}{\uHz}$ (\eqref{eqn:mdl:24}). This value is in exact agreement with $\numaxsun = 3090\pm30\,\uHz$ determined by \citet{Huber2011} who averaged 111 30-day subsets of the same data spanning from 1996 to 2005. It is also in close agreement with $\numaxsun = 3078\pm13\,\uHz$ measured by  \citet{Lund2017} using a weighted average of the green and red VIRGO channel to better match the \Kepler bandpass for a 1150-day timestring with low activity level on the Sun (M. Lund, private communication). The most different measurement we have be able to find in the literature is the measurement of \citet{Gaulme2016} who find $\numaxsun = 3160\pm10\,\uHz$ from a 49 day long VIRGO green channel time series. This suggests that a stable reference value is found as long as one uses a long time series.

We obtained our reference value for $\Dnu_\sun=134.9\pm0.06\,\uHz$ by fitting the the solar frequencies from BiSON \citep{Broomhall2009}, but using only $\ell=0$ modes in the 6 central orders lying closest to the estimated $\numax$ to mimic the procedure we apply to the the giants later. 
This value is in close agreement with $\Dnu_\sun=135.1\pm0.1\,\uHz$ \citep{Huber2011}, $\Dnu_\sun=134.91\pm0.02\,\uHz$ \citep{Lund2017}, and even $\Dnu_\sun=134.82\pm0.08\,\uHz$ from the short 49 day time series mentioned above \citep{Gaulme2016}.

Potential errors in our solar reference values for $\numax$ and $\Dnu$ cause potential errors in mass and radius estimates from the asteroseismic scaling relations. If we take the largest differences between our values and those in the literature as estimates of the largest potential systematic error, these are of 6.9\% and 2.3\% from $\numax$, and 0.3\% and 0.2\% from $\Dnu$, for mass and radius respectively.

\section{Temperatures, radii, masses and RGB mass-loss}\label{sec:stellar_properties}
The fundamental stellar properties mass and radius can be estimated using the so-called asteroseismic scaling relations \citep{Brown1991,ScalingRelations}:
\begin{align}
	\frac{R}{\Rsun} &\simeq \p*{\frac{\Dnu}{\Dnu_\sun}}^{-2} \p*{\frac{\numax}{\numaxsun}} \p*{\frac{\Teff}{\Teffsun}}^{1/2} \label{eqn:radius}\\
	\frac{M}{\Msun} &\simeq \p*{\frac{\Dnu}{\Dnu_\sun}}^{-4} \p*{\frac{\numax}{\numaxsun}}^3 \p*{\frac{\Teff}{\Teffsun}}^{3/2} \label{eqn:mass_1}
\end{align}
In the case of star clusters where we can obtain an independent distance estimate, and hence the luminosity of the stars $L$, the mass-scaling can also be written in three other ways \citep{Miglio2012}:
\begin{align}
	\frac{M}{\Msun} &\simeq \p*{\frac{\Dnu}{\Dnu_\sun}}^{2} \p*{\frac{L}{\Lsun}}^{3/2} \p*{\frac{\Teff}{\Teffsun}}^{-6} \, , \label{eqn:mass_2}\\
	\frac{M}{\Msun} &\simeq \p*{\frac{\numax}{\numaxsun}} \p*{\frac{L}{\Lsun}} \p*{\frac{\Teff}{\Teffsun}}^{-7/2} \, , \label{eqn:mass_3}\\
	\frac{M}{\Msun} &\simeq \p*{\frac{\Dnu}{\Dnu_\sun}}^{12/5} \p*{\frac{\numax}{\numaxsun}}^{-14/5} \p*{\frac{L}{\Lsun}}^{3/10} \label{eqn:mass_4} \, .
\end{align}

To utilize these equations the effective temperatures and bolometric corrections for the stars are needed and for this we employed the calibration by \citet{CV2014}. We used $V$ magnitudes from \citet{Milliman2014} except for two stars that were not in that catalogue and therefore obtained from \citet{Hole2009} (see caption of \tref{tab:table3}). Combining these with $K_s$ magnitudes from the 2MASS catalogue \citep{2MASS} we calculated the $V-K_s$ colour-index and adopted a \emph{nominal} reddening value of $E(B-V)=0.15$ to get bolometric corrections and intrinsic colours using codes and tables from \citet{CV2014}. The reddening choice was made such that the \Teff obtained from the $V-K_s$ colour for the star KIC~5024327 is in exact agreement with a preliminary measurement from high resolution spectroscopy employing the asteroseismic \logg value, \SI{4790}{K}, (Slumstrup et al., in prep.) yielding also $\FeH=+0.02\pm0.10$ which we adopted in the colour-temperature transformations. The reddening value so obtained is identical to that derived by \citep{Bragaglia2001}.

The procedure was done iteratively, making use of \logg from the asteroseismic scaling relations as input for the transformations. We tested whether employing the differential reddening map of \citet{Platais2013} would improve our results, but found that this did not decrease the scatter in our measured masses and distance moduli, suggesting that the precision of the map is not high enough to improve relative \Teff estimates for the stars in our sample and we therefore neglected potential differential reddening effects. We note that our later assumption that all the stars should have the same \emph{apparent} distance modulus is also better this way, since any real differences in the apparent distance modulus caused by differential interstellar absorption effects will be compensated by a slightly offset effective temperature due to differential interstellar reddening.

These choices result in our temperature scale which is on average \SI{34}{K} hotter than the photometric scale of \citet{Lee-Brown2015} based on $B-V$ from several sources and \SI{15}{K} hotter than the IRFM scale of \citet{Casagrande2014}, in both cases with an RMS scatter of \SI{35}{K}. Interestingly, if we adopt the $b-y$ colours from \citet{Casagrande2014} along with the calibration of \citet{Ramirez2005} and their $E(b-y)=0.74{\cdot}E(B-V)$ we obtain a temperature scale which is instead hotter than ours by \SI{48}{K} with an RMS scatter of \SI{50}{K}, partly due to the lower reddening value adopted by \citet{Casagrande2014} compared to us.
The star KIC~5112491 is \SI{33}{K} cooler on our scale than obtained by \citet{Bragaglia2001} using high-resolution spectroscopy (star 333) in a study that found a \logg value only \SI{0.06}{dex} higher than our asteroseismic measurement. KIC~5113061 is \SI{34}{K} cooler on our scale compared to \citet{Thygesen2012} who used an asteroseismic \logg value almost identical to ours to derive $\FeH=+0.01$, very close to our adopted value mentioned above. Our \Teff scale is on average \SI{55}{K} hotter than the one from spectroscopic measurements by APOKASC \citep{Pinsonneault2014} for a subsample of the stars, with an RMS scatter of \SI{62}{K}, when adopting their corrected ASPCAP temperatures. Their uncorrected ASCPCAP temperatures are on average \SI{85}{K} cooler than ours. 

Based on the very good agreement of our \Teff values with the other studies using various methods and calibrations we adopt an uncertainty of 50 K. While it is impossible to prove that this is true in an absolute sense, our later analysis certainly suggests that the estimate is conservative in a relative sense, since otherwise the scatter would be much larger in \fref{fig:corr_dnu_individual}.

\subsection{The distance modulus}
For comparison purposes, the distance modulus to NGC~6819 and the mass of the RGB stars of the cluster can be measured from detached eclipsing binaries as it was done by \citet{Jeffries2013} and \citet{Sandquist2013}. These distance measurements were redone by \citet{Brogaard2015} using the bolometric corrections from \citet{CV2014} which we also employed for the giant stars in the present study. \citet{Brewer2016} added measurements of a third eclipsing cluster member and improved the analysis of the other two. The effective temperatures of the eclipsing components, which is currently determined from photometry, is the main uncertainty on the distance. Therefore, in order to put their distance modulus on the same scale as ours, we reestimated the effective temperature of the binary component WOCS~23009A using the same reddening and colour-temperature relations as for the giants. Since eclipse photometry does not exist for $K_s$ the binary component colours could only be estimated for this one system where the secondary is so faint that the colours of the combined system is effectively the same as those of the primary component. We used the average result from $B-V$, $V-I$, and $V-K_s$ and obtained $(m-M)_V=12.42\pm0.07$ where the uncertainty is taken from the range of results from each individual colour.

\begin{figure}
	\centering
	\includegraphics[width=\columnwidth]{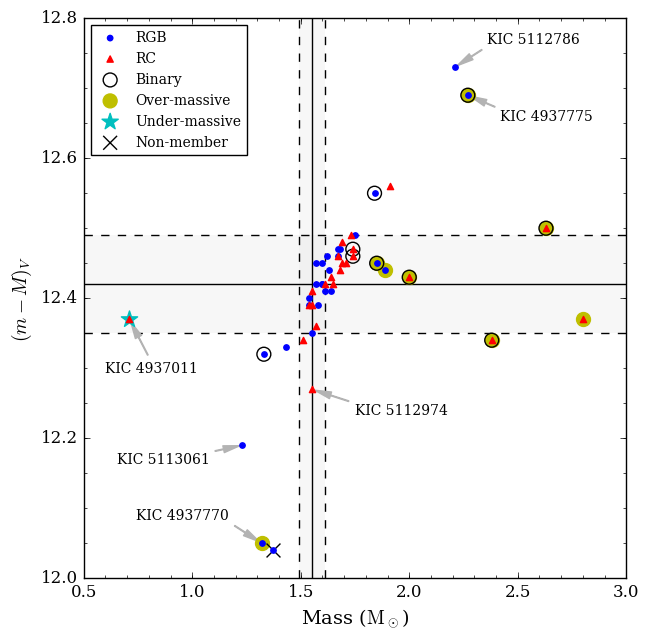}
	\caption{Distance modulus and mass for the cluster stars. In this plot a global correction of 2.54\% was applied to \Dnu. Shaded regions denote the 1-$\sigma$ intervals on the distance modulus and cluster mass (see text for details).}
	\label{fig:distancemodulus_mass}
\end{figure}

\fref{fig:distancemodulus_mass} shows the classical measures of apparent distance modulus, $(m-M)_V=12.42\pm0.07$, and RGB mass \citep{Brogaard2015,Sandquist2013}, $M_\mathrm{RGB}=1.55\pm0.06\,\Msun$ at the $V$-magnitude level of the RC and their 1-$\sigma$ uncertainties (solid and dashed lines) along with the asteroseismic measurements of the individual giant stars, calculated using the scaling relations (\eqref{eqn:radius} and \eqref{eqn:mass_1}). The distance modulus of each giant is calculated using Eq.~(10) in \citet{Torres2010} with bolometric corrections from \citet{CV2014}. 

Alternatively, we used also the surface-brightness-to-angular-diameter calibration of \citet{DiBenedetto2005} to calculate the apparent distance modulus. This calibration is for $(V-K)$ to which we transformed our colours from $(V-K_s)$ using \citet{Carpenter2001}. This allows an interesting comparison, since the effective temperature only appears indirectly, and as a square-root term in the \citet{DiBenedetto2005} calibration, whereas is goes as the fourth power in the luminosity term using \citet{Torres2010}. The latter approach gave distance moduli consistently lower by \SI{0.06}{mag} compared to the former for most of the giants while a completely self-consistent value of $(m-M)_V=12.47$ for the eclipsing component \object{WOCS 23009A} when comparing results from the two methods using $(V-K_s)$ (which is the reasonable thing to do, since systematic errors in colour should then affect each calibration similarly). At this level it is very difficult to identify the cause of discrepancy, since decreasing the giant effective temperatures by just \SI{30}{K} in the former or increasing $E(B-V)$ by only 0.015 in the latter method would bring the distance moduli into exact agreement. Any small systematic error in either method or calibration of the photometry could contribute to the cause.

Most of the measurements in \fref{fig:distancemodulus_mass} are located along a tilted line because because the structure of the asteroseismic scaling relations is such that a measurement error that results in an overestimate of mass also results in an overestimate of radius and through that an overestimate of distance modulus. Stars are excluded from the calculation of mean values, if they have inferred distances deviating by more than 0.2 mag from the ensemble mean (e.g. KIC~5112786, see \fref{fig:distancemodulus_mass}), are binaries, are not cluster members, or represent cluster stars that went through non-standard evolution \citep{Corsaro2012,Brogaard2012,Brogaard2015}. We discuss these stars in detail in \sref{sec:nonstandard}.

\subsection[Corrections to Delta\_nu]{Corrections to $\Dnu$}
\begin{figure}
	\centering
	\subfloat[\label{fig:hr_corr_dnu}]{\includegraphics[width=\columnwidth]{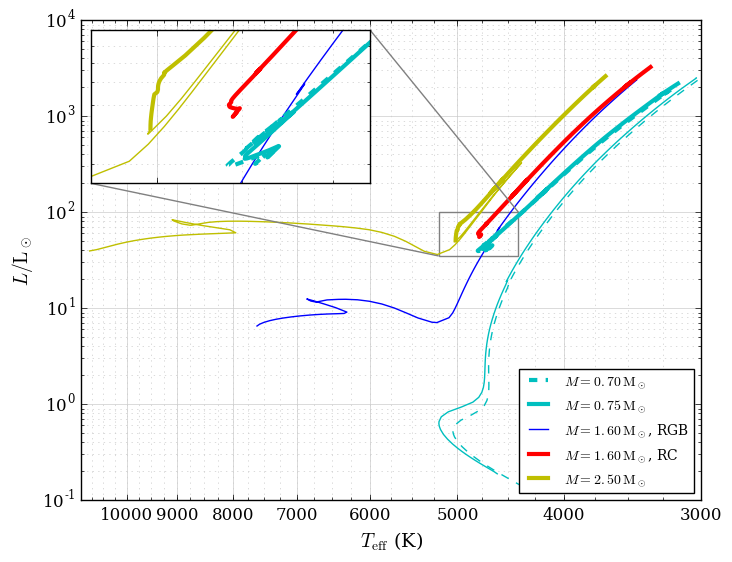}}\\%
	\subfloat[\label{fig:corr_dnu}]{\includegraphics[width=\columnwidth]{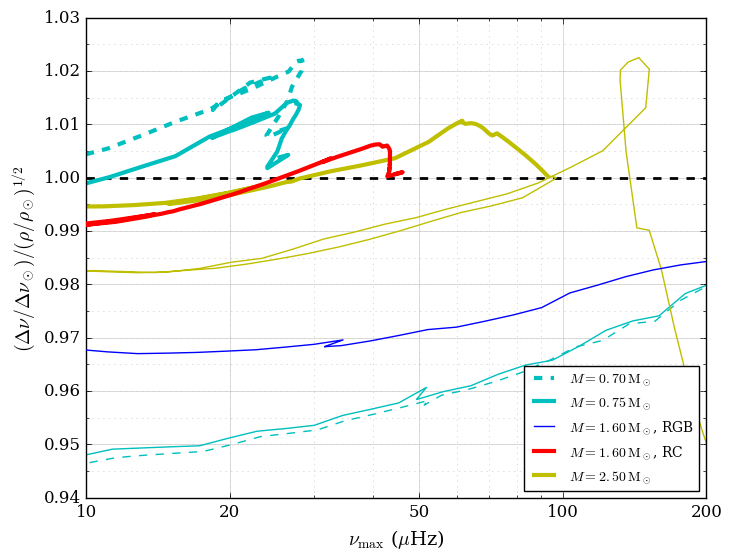}}
	\caption{Stellar model evolutionary tracks for varying mass and $\FeH\!=\!0$. \protect\subref{fig:hr_corr_dnu}: Hertzsprung-Russell  diagram. The insert shows the location of the RC. \protect\subref{fig:corr_dnu}: Deviations from density-scaling for stellar models of selected masses.}
	\label{fig:corr}
\end{figure}

\citet{Miglio2012} found that the asteroseismic scaling relation of $\Dnu$ to mean density is different for helium-burning RC stars compared to RGB stars due to a significantly different structure. Corrections to $\Dnu$ as a function of $\Teff$ for RGB stars were suggested from model calculations by \citet{White2011}. \citet{Miglio2013} extended such calculations all the way to the RC phase. In \fref{fig:distancemodulus_mass}, an empirical positive correction of 2.54\% was applied to $\Dnu$ of all the RGB stars. With this proposed correction to $\Dnu$ the mean distance moduli are identical for the groups of RGB and RC stars, respectively.
Without the correction, the mean distance modulus to the RGB stars is larger by $\Delta DM=0.11\pm0.02$ mag compared to that of the RC stars, which is very unlikely to be true since the two groups of stars belong to the same cluster. With the applied correction the mean distance modulus and mean mass of RGB stars is then also in satisfactory agreement with the measurements from the binary stars at the turn-off and main-sequence as seen in \fref{fig:distancemodulus_mass}. The relative correction between RGB and RC stars is in excellent agreement with predictions in \citet{Miglio2013}. However, those model predictions suggested that about half of the correction should be to RC stars in the negative direction. Our updated theoretically based corrections shown in \fref{fig:corr_dnu} are different in that respect, being in very good agreement with the empirically derived result, also in an absolute sense. We therefore describe those corrections in some detail here while referring the reader to Rodrigues \etal (in preparation) for a full description.

To quantify how well the average large separation of radial modes adheres to the scaling with stellar mean density ($\rho/\rho_\sun\simeq(\Dnu/\Dnu_\sun)^2$), we calculated individual radial-mode frequencies for a set of stellar models ($M=0.75$, 1.60, 2.50\,\Msun and $\FeH=0$), using MESA \citep{Paxton2013}, from the zero-age main sequence to the first thermal pulse (see \citet{Bossini2015} and Rodrigues \etal (in preparation) for an exhaustive description of the models).
Observational measurements of the average \Dnu are limited by the number of frequencies identified around \numax and their uncertainties. Therefore, we must define  a model-predicted $\mean{\Dnu}$ which is closest to how $\mean{\Dnu}$ is estimated from the data. Thanks to the measured individual radial-mode frequencies (see \sref{sec:peakbag}), we were able to construct the average \Dnu from the models in a way that is consistent with that derived from the data.
%
We defined $\mean{\Dnu}$ applying a linear fit as a function of $n$ with weights defined as Gaussian function centred on $\numax$ and of $\mathrm{FWHM}=0.66\,\numax^{0.88}$ \citep{Mosser2012a}. We checked that this definition of $\mean{\Dnu}$ is within 1-$\sigma$ of $\Dnu_\mathrm{fit}$.

Alternative definitions of $\mean{\Dnu}$ can be obtained by taking a weighted average of $\Dnu(\nu)$ itself. This however may lead to significant offsets (especially at low \numax) depending on the reference frequency chosen to apply the weights to. We find the smallest offset if one associates to $\Dnu(\nu)=\nu_{n+1}-\nu_n$ the midpoint frequency between the $\nu_n$ and $\nu_{n+1}$. This explains the difference at low \numax with the corrections presented in \citet{Miglio2013}, where $\nu_n$ was chosen as at the reference frequency.

As is well known current models suffer from an inaccurate description of near-surface layers leading to a mismatch between theoretically predicted and observed oscillation frequencies. These so-called surface effects have a sizeable impact also on the large frequency separation, and on its average value. When utilising model-predicted $\Dnu$ it is therefore important to correct for such effects. As usually done, a first attempt at correcting is to use the Sun as a reference, hence by normalising the $\Dnu$ of a solar-calibrated model with the observed one.
We have calibrated a solar model using the same input physics used in the tracks presented in \fref{fig:hr_corr_dnu} (see Rodrigues et al. in preparation for more details). By comparing the large frequency separation of our calibrated solar model and that from observed solar oscillation frequencies \citep{Broomhall2014} we find that the predicted average ($\Dnu_\mathrm{mod}=\SI{136.1}{\uHz}$) is a 0.8\% larger than the observed one ($\Dnu_\mathrm{obs}=\SI{134.9}{\uHz}$). We have therefore normalised the model predicted $\mean{\Dnu}$ to that of the calibrated solar model, leading to a model-suggested correction to $\Dnu$ as shown in \fref{fig:corr_dnu} for stars positioned in the Hertzsprung-Russell diagram according to \fref{fig:hr_corr_dnu}.

\begin{figure}
	\centering
	\includegraphics[width=\columnwidth]{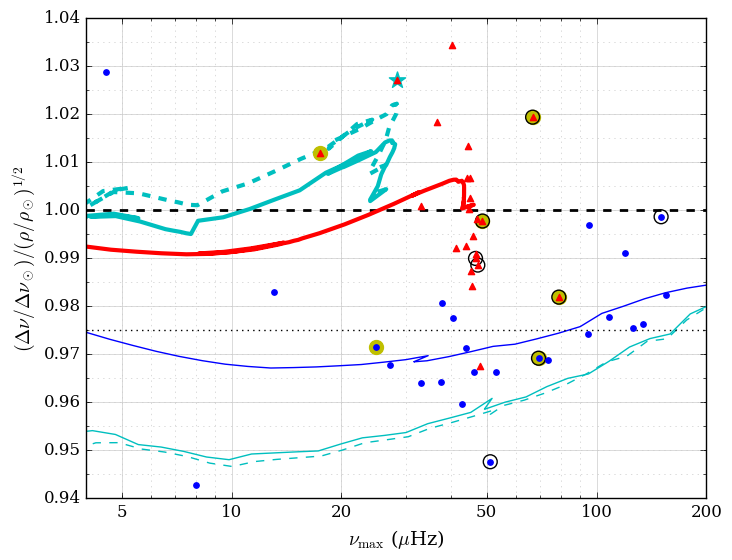}
	\caption{Comparison between theoretical and empirical corrections to scaling relations.}
	\label{fig:corr_dnu_individual}
\end{figure}

In \fref{fig:corr_dnu_individual} we compare these theoretical corrections to empirical ones calculated from our measurements by assuming the distance modulus to be $(m-M)_V = 12.42$ and on a star-by-star basis adjusting \Dnu until mass eqns. (16) and (17) (and therefore all four mass equations) yield the same mass. As seen, there is a general very good agreement between the theoretical and empirical correction with a scatter caused by errors in the measurements of \Dnu, \numax, and \Teff. The agreement strongly supports the theoretical corrections to \Dnu. This also suggests that no correction is needed for the other global parameter \numax (at least not at this metallicity). The fact that we are able, for the first time, to observationally confirm that the size of the \Dnu correction changes with evolution up the RGB also shows that there is no significant evolutionary state dependent correction to \numax on the RGB. The agreement on the \Dnu correction for both the RGB and RC phases also confirms that \numax should remain uncorrected, also for the RC phase of evolution.

By comparing the measured position relative to the theoretical prediction in \fref{fig:corr_dnu_individual}, one can determine whether the star is on the RGB, in the RC, or an even later stage in evolution. This has important implications for the future exploitaion of asteroseismology of giant stars, since all that is needed is the global asteroseismic parameters \Dnu and \numax and a distance measurement. With the latter expected from Gaia \citep{Perryman2001} in the near future, this opens up the possibility to know the evolutionary state of stars without measuring the more difficult and observationally demanding gravity mode period spacing \citep{Bedding2011}, and thus it can be done for many more stars with less effort. This is also potentially very important for upcoming missions like TESS \citep{Ricker2014}, where individual frequencies can not be obtained for large fractions of the observed stars due to the limited time-coverage. Only one of the regular stars in NGC~6819 would be clearly misclassfied according to \fref{fig:corr_dnu_individual} (the RC star which falls on the RGB relation just below 50 $\uHz$) while a few would be uncertain due to measurement errors. However, the identification becomes easier for stars of lower mass, since the \Dnu correction varies more at lower masses. This allows a very clear classification as a RC star for an undermassive cluster member KIC~4937011, marked with a cyan star in \fref{fig:corr_dnu_individual} and described in detail in \sref{sec:li}.
We emphasize the potential of discriminating between the RGB, RC and AGB phases using only the asteroseismic parameters \Dnu and \numax for any star with a known distance once future investigations verify that we can trust the correction to \Dnu and that no correction to \numax is needed for all masses and metallicities that we are probing. Alternatively, for stars with a known distance where we can determine the evolutionary state independently, we can exploit the fact that we know the \Dnu correction to obtain \Teff without the use of colours or spectra.

\subsection{Mass loss}
The mean masses of the RGB stars and RC stars in NGC~6819 were found to be $1.61\pm0.02\,\Msun$ and $1.64\pm0.02\,\Msun$, respectively, independent of whether empirical or theoretical corrections to \Dnu were used. A systematic uncertainty of at least the same level could be present, since changing $E(B-V)$ in our procedure by \SI{0.01}{mag} (corresponding to about \SI{25}{K}) changes the mean masses by $0.01\,\Msun$. While such systematic uncertainties on e.g. effective temperatures add to the uncertainties of the absolute masses, the difference $\Delta M=-0.03\pm0.01$ is almost insensitive to systematics. We note in particular that we obtain the exact same value for $\Delta M$ when calculated using either of the Eqns.~(\ref{eqn:mass_1})--(\ref{eqn:mass_4}), one of which does not include $\Dnu$ and therefore independent of potential uncertainties in the correction of this parameter.
This confirms the result of \citet{Miglio2012} of $\Delta M=-0.03\pm0.06$ between RGB and RC, but now with much higher confidence. A commonly used mass-loss law is that of \citet{Reimers1975}:
\begin{equation}
	\frac{d M}{d t} = 4\times10^{-13} \eta_\mathrm{R} \frac{L R}{M} \, ,
\end{equation}
where $L$, $R$, and $M$ are the stellar luminosity, radius and mass in solar units, and $\eta_\mathrm{R}$ is a constant with units $\Msun$ yr$^{-1}$. Isochrone comparisons to our measured $\Delta M$ like in Fig. 7 of \citet{Miglio2012} suggest $\eta_\mathrm{R}$ in the range 0 to 0.1 with a preference for 0. This is much lower than $\eta_\mathrm{R}=1.4$ suggested by the calibration by \citet{KudritzkiReimers1978} and also significantly lower than the value of $\eta_\mathrm{R}$=0.4 often adopted for globular clusters \citep[see \eg][]{RenziniFusiPecci1988}.

\citet{Schroder2005} derived a more physically motivated mass-loss formula:
\begin{equation}
	\frac{d M}{d t} = \eta_\mathrm{SC} \frac{L R}{M} \p[\Big]{\frac{\Teff}{\SI{4000}{K}}}^{3.5} \p[\Big]{1+\frac{g_{\sun}}{4300\times g_{\star}}} \, ,
\end{equation}
with $L$, $R$, and $M$ defined as before, $\Teff$ and $g_{\star}$ the effective temperature and surface gravity of the star and $g_{\sun}$ the surface gravity of the Sun. In this case, $\eta_\mathrm{SC}=8\times10^{-14}$ $\Msun$ yr$^{-1}$. 
This is in conflict with our measurement since, without a recalibration of the constant $\eta_\mathrm{SC}$ in their relation, it predicts about the same mass loss for NGC~6819 as Reimers with $\eta_\mathrm{R}=0.4$. However, such a recalibration would cause the relation to be in conflict with more direct measurements of mass-loss rates which match well the relation with the current value of $\eta_\mathrm{SC}$ \citep{Schroder2007}.

It is interesting to note that if the RGB mass-loss in NGC~6791 also remains at the same value as derived by \citet{Miglio2012} when rederived with higher precision through the same procedure as ours for NGC~6819, then results on the two clusters NGC~6791 and NGC~6819 together suggest a much stronger mass dependence for the RGB mass-loss than suggested by the current formulations.

The measured masses of both RGB and RC stars are matched extremely well by a PARSEC isochrone of solar metallicity, no RGB mass-loss and an age of \SI{2.25}{Gyr} --- an age also in close agreement with that derived from eclipsing binary stars by \citet{Brewer2016} and from the white dwarf cooling sequence by \citet{Bedin2015}.

\section{Stars that experienced non-standard evolution}\label{sec:nonstandard}
We attribute a membership classification to the stars based on the radial-velocity (RV) and proper-motion (PM) study by \citet{Milliman2014} combined with our asteroseismic distance and mass measurements.
In \tref{tab:table2} a full list of classifications of the analysed stars is given. With regards to the nomenclature on cluster membership we use the following extended convention: `S' and `B' respectively denotes single and binary stars, `O' and `U' denoting over- or under-massive and `M' or `N' denoting member or non-member of NGC~6819. As an example, `BOM' would denote a binary over-massive star that is a cluster member.

\subsection{KIC~4937011 -- a Li-rich low mass red clump member}\label{sec:li}
The star KIC~4937011 has a 90\% proper-motion probability of being a cluster member according to \citet{Sanders1972} and was classified as a radial-velocity cluster member by \citet{Hole2009}. However, an asteroseismic analysis by \citet{Stello2011} classified the star as a non-member. \citet{Twarog2013} found this star to be lithium-rich and argued that it could be a cluster member red giant below the RGB bump and that the asteroseismic mis-match found by \citet{Stello2011} might be due to the star being in a special evolutionary stage.
Recently, \citet{Milliman2014} published an updated radial velocity membership study of NGC~6819 that also contains the proper-motion membership information from \citet{Platais2013}. While in that study KIC~4937011 remains a radial velocity member, the proper motion membership is 0\%, in great contrast to the result from \citet{Sanders1972}, causing additional confusion about the cluster membership. \citet{Carlberg2015} investigated KIC~4937011 in greater detail with their evidence pointing strongly towards cluster membership.

Our asteroseismic analysis revealed that KIC~4937011 is a RC star according to the mixed-mode period spacing \citep{Bedding2011}. However, in accordance with \citet{Carlberg2015} we found KIC~4937011 to have a low-mass of $0.71\pm0.08\,\Msun$, much lower than the other RC stars in the cluster. Given a negative correction to $\Dnu$ of 1.5\% relative to the other RC stars, as suggested by \fref{fig:corr_dnu} and \citet{Miglio2013} for such low-mass RC stars, the star has self-consistent mass estimates from the four different mass-equations above for a star located at the cluster distance (See cyan star in \fref{fig:distancemodulus_mass}). While this does not prove that the star is a cluster member, it is strongly suggestive. The position of the star in the CMD of \fref{fig:hrd} is also consistent with that of a RC star with a much lower mass than the other RC stars in the cluster, which we have confirmed by comparing to ZAHB loci in \citet{Rosvick1998}. Although not evident from the publication, we have checked that a RC star with a mass of $0.71\pm0.08\,\Msun$ would be located very near but a bit fainter than the reddest point of the ZAHBs they presented (D.~A.~VandenBerg, private communication), in excellent agreement with the relative magnitude and colour between KIC~4937011 and the other RC stars, and the predicted relative positions in the HR diagram of \fref{fig:hr_corr_dnu}. Since the star also has both \FeH and $[\alpha/\mathrm{Fe}]$ very similar to the other cluster stars according to measurements from the APOGEE survey \citep{Carlberg2015}, it seems very likely that KIC~4937011 is a cluster member in the RC phase of evolution that for some reason experienced enhanced mass-loss. Since the star was found to be a lithium-rich giant by \citet{Twarog2013}, this might indicate that the lithium production is related to the high mass-loss, a clue that might help solve the general puzzle of lithium-rich giants, as discussed \eg by \citet{SilvaAguirre2014}. KIC~4937011 is to the best of our knowledge only the third lithium-rich giant to be examined by asteroseismology\footnote{The other two being KIC~9821622 \citep{Jofre2015} and KIC~5000307 \citep{SilvaAguirre2014}.}, and the second one to be confirmed as a RC star. 
The strong indication that this star experienced higher-than-average mass-loss was only possible because it can be compared to other cluster members and thus demonstrates the importance of star clusters as asteroseismic science labs. It also highlights the caution that needs to be taken when age-dating RC stars in the field with asteroseismology, since some of them might have experienced higher-than-average mass-loss.

As an alternative to a higher than average mass-loss for KIC~4937011, the star could be a normal field RC giant that just happens to be located within the cluster. A repeat measurement using new observations of the proper-motion would be useful to determine whether it also moves with the cluster. In any case, the likelihood of having a chance overlap of an extremely old (as implied by the mass if mass-loss was normal; The age from the SAGA Survey is $14.7\pm2$\,Gyr neglecting mass-loss \citep{Casagrande2014}) solar-metallicity giant and a solar-metallicity cluster of much younger age seems very small.

\subsection{Cluster members with above average mass}
As seen in \fref{fig:distancemodulus_mass} there are a number of stars located at the cluster distance but with a higher mass than most of the stars. One cannot just dismiss these as non-members since, in addition to being at the cluster distance, they are also found to also be both PM and RV members in the study by \citet{Milliman2014}, many of them as long period SB1 binary members suggesting a mass-transfer origin for the higher mass. 

We find a total of 6 overmassive giant members in our sample of 51 cluster members. While this is not a complete sample it strongly indicates that more than 10\% of the giants in NGC~6819 are overmassive compared to the ensemble. We note that there are even more suspected overmassive candidates in the so-called superstamps observed in the central cluster region by \Kepler but not yet exploited for this purpose.
Our interpretation of these over-massive giants is that they underwent non-standard evolution through mass-transfer in a blue-straggler phase. This prevents estimates of age from their masses. Whether or not a similar number of over-massive giants is to be expected for field stars depends on whether binary fractions are significantly different in the field compared to open clusters and whether NGC~6819 can be considered a typical cluster in this respect. An additional complication is that most of the over-massive stars identified in NGC~6819 belong to very long period binary systems. Therefore it is not clear how the percentages will change for samples of stars where binarity can be ruled out. The potential appearance of a significant fraction of over-massive stars arising from non-standard evolution is certainly something that should be kept in mind when deriving asteroseismic ages for giant stars, especially when finding younger stars than expected. A good example of this is the `Young $\alpha$-rich stars' found by \citet{Chiappini2015} and \citet{Martig2015} which are now thought to be the result of mass-transfer rather than young age due to follow-up spectroscopy identifying many of them as long-period binaries \citep{Yong2016,Jofre2016}, exactly like the majority of the overmassive stars in our sample.

We discuss the individual over-massive giants below, referring to measured and derived values in tables \ref{tab:table1}--\ref{tab:table3}.

\subsubsection*{KIC~5024272}
This star is a high probability PM and RV single member \citep{Milliman2014} with an asteroseismic distance consistent with the cluster but a mass of about 2.8 \Msun, much larger than the average RGB mass of $1.61\pm0.02$. This suggests that the giant formed from two stars that merged to form a massive single star. The single member status determined by \citep{Milliman2014} is based on a total of 8 radial velocity measurements from two different telescopes, 4 epochs at each location covering well separated intervals of about 1000 days, and with about 5000 days between observations at the two telescopes. Still, due to measurement uncertainties and the unknown inclination of a potential binary orbit, a binary companion cannot be completely ruled out without further investigation. If a companion exists, it would have a mass below 0.4 \Msun (twice the turn-off mass minus the present mass of the overmassive star) if the system formed from a binary.
 
The CMD location about \SI{1.5}{mag} brighter in $V$ than the RC (brightest star in \fref{fig:hrd}) is consistent with either the RGB or the AGB evolutionary phase. By making use of the four mass equations and relying on the predicted correction to \Dnu for the RGB and AGB phase (\fref{fig:corr_dnu}), respectively, for a star of this approximate mass we find a slight preference for the AGB phase, since that requires the smallest correction to reddening (and thus \Teff) or \numax in order for all four mass equations to yield the same mass. If we assume that it is \Teff which needs a correction due to e.g. differential reddening or errors in the colour measurement then \Teff needs to be higher than our measurement by about 30 K for the AGB phase scenario and about 60 K higher for the RGB scenario. This suggests that if anything our \Teff is underestimated. This would also explain why \citet{Lee-Brown2015} who used a \Teff which is 145 K lower than our measurement obtained a low value of $\FeH=-0.37$ since a higher \Teff would also result in a higher \FeH more consistent with the average cluster metallicity. 

\subsubsection*{KIC~5023953}
This is another high probability PM and RV member, this time of SB1 type \citep{Milliman2014}. We classified this star as a RC star from the observed period spacing and its location in the RC in the CMD, which is also supported by the self-consistency of the mass from the four equations without applying a correction to \Dnu (See \fref{fig:corr_dnu}). It has very similar values of $\Dnu$ and $\epsilon_{c}$ compared to the other RC stars, but it has a higher mass, $M=1.83\pm0.10\,\Msun$. This only agrees with expectations if the star is in a late core-helium-burning phase \citep[see Fig. 4 in][]{Stello2013}.

\subsubsection*{KIC~5112880}
Our analysis shows this to be an overmassive RGB star at cluster distance with a self-consistent asteroseismic mass of $M = 1.90\pm0.14\,\Msun$, in agreement with the single member status of \citet{Milliman2014}.

\subsubsection*{KIC~5112361}
This is an overmassive star at the cluster distance with $M = 1.85\pm0.06\,\Msun$ and a SB1 binary member according to \citep{Milliman2014}. The star is mentioned as outlier A in the $\Dnu-\Delta P$ diagram by \citet{Corsaro2012}. However, our estimate of the observed period spacing is close to 60 seconds and thus we see no disagreement with either the theoretical or empirical prediction for the RGB phase \citep[see Fig.~4 in][]{Stello2013}. The correction to \Dnu (\fref{fig:corr_dnu}) also supports the RGB phase, since assuming RC for the correction results in discrepant masses from the four equations and a too large distance modulus.

\subsubsection*{KIC~5024476}
With a mass close to $M = \SI{2.38}{\Msun}$ this giant is overmassive and also an SB1 binary member \citep{Milliman2014}.
It is mentioned as outlier D in the $\Dnu-\Delta P$ diagram by \citet{Corsaro2012}. In general it is more difficult to use the $\Dnu-\Delta P$ diagram to distinguish between RGB and helium-burning phases for such high mass stars because the RGB phase of high mass stars also displays and increased period spacing \citep[see Fig.~4 in][]{Stello2013}. However, in this case, the value found by \cite{Corsaro2012} is high enough that a core-helium burning secondary clump (2C) phase can be inferred when combined with the high mass. To support this further, we investigated the mass self-consistency using the theoretical correction to \Dnu according to \fref{fig:corr_dnu} assuming either an RGB or 2C scenario.

For both cases a correction to \Teff was also needed, +27 K for the secondary clump (2C) phase, +102 for RGB. The \Teff correction for the RGB case is at the $2\sigma$ level. Additionally, the resulting \Teff for the RGB scenario is almost 200 K hotter than the RGB evolutionary track for $M = 2.38\,\Msun$ at the measured luminosity, whereas for the 2C scenario the \Teff is only \SI{28}{K} cooler than the model prediction from \fref{fig:hr_corr_dnu}. We take this as further confirmation that the star is in the 2C phase.

We derived the minimum mass of the secondary component from the semi-amplitude of the spectroscopic orbit from \citet{Milliman2014} and the asteroseismic mass of the giant component. If we assume that the system originates from two normal stars in the cluster then we can also estimate the maximum mass of the secondary component as the difference between the mass of the giant and two stars of the turn-off mass at about the time when the two stars started the mass transfer $(2\times1.61-2.38),\Msun=0.84\,\Msun$. Since the minimum mass turns out to be as high as \SI{0.6}{\Msun} we have the absolute mass of the secondary component constrained to be 0.6--0.84$\,\Msun$ consistent with the mass of a white dwarf in NGC~6819 according to \citet{Bedin2015}. 

The distance modulus for KIC~5024476 is different by \SI{0.44}{mag} between the two distance measures we use, which could be indicating that magnitudes and/or colours are affected by a secondary component. However, a white dwarf, which seems the only likely secondary star due to the need of significant mass-transfer and the secondary mass derived, would either be too faint to affect colours or cause and overestimate of \Teff of the giant if the white dwarf is still in an early hot phase. Our above investigation required a hotter \Teff, not lower. We therefore take the discrepant value of the distance modulus from the surface-brightness method as a further indication that the giant is in the 2C and that the surface-brightness-to-angular-diameter calibration does not hold for massive core-helium burning giants.

\subsubsection*{KIC~5024414}
This star is mentioned as outlier C in the $\Dnu-\Delta P$ diagram by \citet{Corsaro2012}. It is an SB1 binary member \citep{Milliman2014} with a mass of roughly $\SI{2.6}{\Msun}$.
As mentioned above it can be difficult to use the $\Dnu-\Delta P$ diagram to distinguish between RGB and secondary clump phases, since for such high mass stars both phases can appear in overlapping regions of the diagram \citep[see Fig.~4 in][]{Stello2013}. However, in this case the observed period spacing is high enough (about \SI{170}{s}) to give strong preference to the 2C scenario for this mass.
The value of $\epsilon_c$ also indicates that the star is in the core-helium burning phase.

The distance modulus for this star is different by \SI{0.57}{mag} between the two distance measures we use, indicating that magnitudes and/or colours could be affected by a secondary component. B01 derived a spectroscopic \Teff of 4740 K which is much lower than our photometric value of 5049 K, which also suggests that colours are not directly related to \Teff for this star. On the other hand, our photometric \Teff is in good agreement with helium burning stars of similar metallicity and mass in NGC~6811 as derived by \citet{Molenda2014}. Also, the theoretical correction to \Dnu requires in this case only relatively small negative corrections \Teff to reach mass self-consistency, \SI{-39}{K} for RGB, \SI{-67}{K} for 2C, with the resulting \Teff most consistent with the model for the 2C scenario. The fact that the measured \Teff seems to be close to the true value allows us to interpret the discrepant distance modulus from the surface-brightness-to-angular-diameter method as being caused by the calibration being invalid for massive core-helium burning giants.

As for KIC~5024476 we can constrain the mass of the secondary binary component which in this case turns out to be in the range 0.47--0.72$\Msun$, again consistent with a white dwarf. 

\subsection{Members and non-members}\label{sec:nonmembers}

\subsubsection*{KIC~5024143}
A single non-member according to \citet{Milliman2014}, based on PRV$=93\%$ and PM$=0\%$. Everything but but proper motion points to membership, so we re-classify to SM. 

\subsubsection*{KIC~4937257}
This is a giant with a mass of $\SI{1.08}{\Msun}$ in the RC evolutionary phase. According to \FeH measurements (see \tref{tab:table3}) it has a metallicity close to solar.
From the distance modulus and CMD position (at $V=14.23, V-K_s=3.02$ in \fref{fig:hrd}) this star is a clear non-member, located behind the cluster. Instead, this star turns out to be very similar to RC stars of another open cluster, NGC~188.
\citet{Meibom2009} measured the turn-off mass of NGC~188 from an eclipsing binary member to be very close to $1.1\,\Msun$. The mass of the RC stars will be close to that of the turn-off mass, since the slightly higher mass of the RGB will almost compensated by the RGB mass-loss before reaching the RC as demonstrated for another old open cluster NGC~6791 by \citet{Brogaard2012}.
By comparing the $V-K_S$ colour of KIC~4937257 to that of KIC~4937011, the low-mass RC star and the colour of the other RC stars in NGC~6819, taking into account their relative masses, we estimated that $E(B-V)$ is higher for KIC~4937257 compared to the NGC\,6819 by about 0.006 magnitude. Applying this to our \Teff derivation, we obtained $\Teff=\SI{4600} {K}$ which is also the preliminary spectroscopic temperature obtained for a RC star in NGC~188 (Slumstrup et al., in prep.).
Employing the CMD of NGC~188 produced by \citet{Stetson2004} we estimate the $V$ magnitude of the RC stars to be \SI{12.4}{mag}.
Adopting the apparent distance modulus from \citet{Meibom2009}, $(V-M_V)=11.24$ results in an absolute $M_V$ magnitude of the RC stars of \SI{1.16}{mag}. For comparison, $M_V$ for KIC~4937257 is \SI{1.07}{mag}, which is consistent when taking into account uncertainties in the photometries and distance moduli of the two clusters. Summarising all this, KIC~4937257 is a RC star located in the field behind NGC~6819 with mass, \Teff, metallicity and absolute magnitude very similar to the RC stars of the old open cluster NGC~188.
By extension, the age of KIC~4937257 should also be very close to that of NGC~188, $6.2\pm0.2$\,Gyr \citep{Meibom2009}.

KIC~4937257 is an example of the rare case when proper motion, radial velocity and metallicity measurements all erroneously point to cluster membership.

\subsubsection*{KIC~5024043}
A non-member according to \citet{Milliman2014}, based on PRV$=94\%$ and PM$=0\%$. \citet{Sanders1972} has PM=65\% which would cause doubts about membership without further information. However, the asteroseismic distance modulus is \SI{0.4}{mag} lower than that of the cluster, the CMD position is in the cluster RC while the asteroseismic classification is RGB with a radius which is also lower than that of the RC cluster stars. The metallicity is also consistently lower than the cluster mean according to the three literature \FeH measurements in \tref{tab:table3}. Based on this evidence we conclude that this is a foreground RGB star with a mass of {$1.32\pm0.04\,\Msun$.

\subsubsection*{KIC~5023889}
The apparent distance modulus $(m-M)_V=11.24$ for this RGB star is much shorter than the cluster distance $(m-M)_V=12.42$ and it is also a single non-member according to \citep{Milliman2014} based on based on PRV$=93\%$ and PM$=0\%$. Shifting the CMD postion of the star in \fref{fig:hrd} with the relative distance moduli puts it right on the RGB sequence where it would be expected to end up if it was a member since the metallicity is about the same as the cluster (\tref{tab:table3}). Therefore, there is no indication that the reddening is significantly lower than the cluster and the \Teff should thus be reliable for the mass estimate of $M=1.40\pm0.05\,\Msun$. 

\subsection*{Uncertain cases}

\subsubsection*{KIC~5112974}
On the mass-distance plane in \fref{fig:distancemodulus_mass} this single member \citep{Milliman2014} RC star seems slightly deviant compared to the other RC stars, displaying a mass of $\SI{1.55}{\Msun}$ and $(m-M)_V=12.27$. However, taking into account that this is an evolved RC star with a \Dnu value lower than the vast majority of RC stars suggests a theoretical \Dnu correction that lessons the tension. Allowing additionally for the \Teff to be 61 K hotter than our measurement would result in complete agreement with the mean mass and distance for the RC stars. We thus attribute the slightly discrepant values for this star to measurement errors.

\subsubsection*{KIC~5113061 and KIC~5112786}
These stars are on the mass-distance slope expected for members with rather large errors in \fref{fig:distancemodulus_mass}. In the CMD they are located on the upper RGB sequence. Consequently, they present low values of \Dnu and \numax, and thus relative uncertainties that are large enough that they are still consistent with cluster membership in agreement with their SM status according to \citet{Milliman2014}.

\subsubsection*{KIC~4937770 and KIC~4937775}
These stars are located bluer than the single member RGB and below the RC in the CMD. Thus, if they are members, they are expected to be overmassive RGB stars. A higher than average RGB mass would also explain the outlier status of KIC~4937770 in \cite{Corsaro2012} (Outlier B) based on the $\Dnu-\Delta P$ diagram. While not considered by Corsaro et al., a higher mass can explain the $\Delta P$ value due to the fact that RGB stars with masses above about $\SI{2}{\Msun}$ do show a gravity mode period spacings higher than that of RGB stars of lower mass \citep[see Fig.~4 in][]{Stello2013}.

However, both stars appear to be significantly off the cluster distance in \fref{fig:distancemodulus_mass}, which complicates their interpretation. Their \Teff from $b-y$ is lower by $\sim$\SI{100}{K} compared to $V-K$ although the average difference for the ensemble is \SI{50}{K} in the other direction, suggesting issues with the photometry. Furthermore, the uncertainties on the asteroseismic parameters are large and \numax is in the range where the specific choice of background fit can change the derived \numax by more than 1 sigma (see \fref{fig:background_comparison}; \numax range 80--110 \uHz), which complicates the analysis further.
The membership study by \citep{Milliman2014} suggests membership for both, KIC~4937775 being a binary member and KIC~4937770 a single member, although with proper motion membership probability of only 16\% for the latter.

\section{Summary, conclusions and outlook}\label{sec:conclusion}
Our extensive peakbagging effort on \Kepler light curves of evolved red giant stars of the open cluster NGC~6819 allowed new insights into both the cluster and asteroseismology.
The fundamental improvement obtained by measuring individual oscillation modes and constructing the global parameters from these, is that exactly the same can be done for stellar models. This means that we are actually comparing the same thing in the measurements and models -- something that is not possible if only the global parameters are available (\eg the power spectrum of the power spectrum). We would therefore argue that this method is much more \emph{accurate}, even if the \emph{precision} would be comparable.

We found a remarkable agreement between empirical and theoretical corrections to the asteroseismic scaling relations for the red giants in NGC~6819. To verify that this holds at other masses and metallicities high precision asteroseismic measurements of giant stars in more open clusters \citet{Brogaard2015} and/or giant stars in binary systems \citep[\eg][]{Frandsen2013,Brogaard2016} should be explored.   

In the cases of the studied 51 red giant stars, we find that there is an excellent agreement between $\Dnu_\mathrm{global}$, obtained by using the PSPS-method, and $\Dnu_\mathrm{fit}$, obtained from individual frequencies (see \fref{fig:lsep_comparison}).
This is of vital importance for the many stars where individual frequencies can not be obtained, as this means that $\Dnu_\mathrm{global}$ can be directly compared to the values from stellar models, be it with larger uncertainties than what can be obtained with individual frequencies.
We do, however, find that a little care has to be taken with respect to \numax and its correlation with the treatment of the granulation noise background.
Differences between \numax obtained from different choices of background models are small, but can introduce differences on the percent-level, which, in the era of precision asteroseismology, could introduce significant contributions to the errors on estimated masses of stars.

We confirmed what was previously shown for stellar models \citep[\eg][]{Montalban2010a,Montalban2010,Montalban2011}, that the small frequency separation, $\dnu{02}$, can be used to distinguish the stellar evolutionary state between RGB and RC stars. Furthermore, we identified a new potential way to also make this distinction based on the empirical correction to \Dnu for stars with a known distance. Although more studies are needed to set the ranges of applicability, this opens promising opportunities for asteroseismology with the upcoming TESS and PLATO missions complemented by Gaia distances.

The mean masses of the RGB stars and RC stars in NGC~6819 are $1.61\pm0.02\,\Msun$ and $1.64\pm0.02\,\Msun$, respectively. While systematic uncertainties on e.g. effective temperatures add to these absolute masses, the difference $\Delta M=-0.03\pm0.01\,\Msun$ is much better defined and almost insensitive to systematics. Assuming a Reimers mass-loss law, model comparisons suggest $\eta_\textup{R}$ in the range 0 to 0.1 with a strong preference for no mass-loss at all.

Some of the stars that are outliers relative to the ensemble are revealed to be overmassive members that likely evolved via mass-transfer in a blue straggler phase, and KIC~4937011 is identified as a low-mass Li-rich cluster member in the RC phase that experienced very high mass-loss during its evolution. Such over- and undermassive stars need to be considered when studying field giants, since the true age of such stars cannot be known and there is currently no way to distinguish them from normal stars. If NGC~6819 can be assumed representative of the general field star population then about 10\% of all giant stars did not evolve a single stars.

Our finding of non-members among stars with high cluster membership probabilities from radial-velocity and proper motion information and four members among stars with proper motion membership of 0\% remind us that these tools are statistical in nature and therefore not to be trusted blindly for individual stars. It seems that for the particular study by \citet{Milliman2014} especially a proper motion membership of 0\% should be considered with caution.

The extracted asteroseismic parameters and individual frequencies for this unique sample of stars will be useful for additional future studies of asteroseismology and stellar evolution.

\section*{Acknowledgements}
Funding for the Stellar Astrophysics Centre is provided by The Danish National Research Foundation (Grant DNRF106). The research was supported by the ASTERISK project (ASTERoseismic Investigations with SONG and Kepler) funded by the European Research Council (Grant agreement no.: 267864).

\appendix
\onecolumn

\section{Tables}

\begin{table}
\centering\tiny
\begin{threeparttable}
	\caption{Extracted asteroseismic parameters.}
	\label{tab:table1}
	\tiny
	\begin{tabular}{crrccrrrrrrrr}
	\toprule
	\mc{KIC} & \mc{WOCS\tnote{1}} & \mc{$\Teff$\tnote{2}} & \mc{Orders\tnote{3}} & \mc{Background} & \mc{$\mean{\dnu{02}}$}  & \mc{$\mean{\Gamma_{n0}}$} & \mc{\numax}    & \mc{$\epsilon_c$\tnote{4}} & \mc{$\Dnu_c$\tnote{4}} & \mc{$\epsilon$} & \mc{$\Dnu$} & \mc{$\Dnu_\mathrm{corr}$\tnote{5}} \\
	         &                    & \mc{(K)}              &                      &                 & \mc{(\uHz)}             & \mc{(\uHz)}               & \mc{(\uHz)}    &                            &  \mc{(\uHz)}           &                 & \mc{(\uHz)} & \mc{(\uHz)}            \\
	\midrule
   4937011 &    7017 & 4669 & 0 & (\ref{eqn:mdl:4})  &                --- &                --- & $  28.34{\pm}0.44$ &                   --- &                --- &                --- & $   4.08{\pm}0.10$ & $   3.98$ \\
   4937056 &    2012 & 4788 & 0 & (\ref{eqn:mdl:24}) &                --- &                --- & $  47.28{\pm}0.62$ &                   --- &                --- &                --- & $   4.76{\pm}0.12$ & $   4.82$ \\
   4937257 &    9015 & 4601 & 5 & (\ref{eqn:mdl:22}) & $   0.57{\pm}0.01$ & $   0.24{\pm}0.01$ & $  34.54{\pm}0.24$ & $      1.02{\pm}0.16$ & $   4.10{\pm}0.10$ & $   0.87{\pm}0.04$ & $   4.18{\pm}0.02$ &       --- \\
   4937576 &    5016 & 4542 & 7 & (\ref{eqn:mdl:24}) & $   0.46{\pm}0.01$ & $   0.10{\pm}0.01$ & $  32.94{\pm}0.26$ & $      0.97{\pm}0.03$ & $   3.54{\pm}0.02$ & $   1.06{\pm}0.06$ & $   3.51{\pm}0.03$ & $   3.64$ \\
   4937770 &    9024 & 4930 & 0 & (\ref{eqn:mdl:22}) &                --- &                --- & $  87.19{\pm}1.29$ &                   --- &                --- &                --- & $   7.96{\pm}0.19$ &       --- \\
   4937775 &    9026 & 5060 & 0 & (\ref{eqn:mdl:2})  &                --- &                --- & $  91.82{\pm}0.75$ &                   --- &                --- &                --- & $   7.30{\pm}0.18$ &       --- \\
   5023732 &    5014 & 4588 & 6 & (\ref{eqn:mdl:4})  & $   0.41{\pm}0.01$ & $   0.10{\pm}0.02$ & $  27.20{\pm}0.26$ & $      0.88{\pm}0.14$ & $   3.12{\pm}0.06$ & $   1.00{\pm}0.08$ & $   3.08{\pm}0.03$ & $   3.19$ \\
   5023845 &    8010 & 4845 & 6 & (\ref{eqn:mdl:2})  & $   1.09{\pm}0.02$ & $   0.11{\pm}0.01$ & $ 108.47{\pm}0.57$ & $      1.22{\pm}0.06$ & $   8.91{\pm}0.05$ & $   1.26{\pm}0.04$ & $   8.87{\pm}0.04$ & $   9.08$ \\
   5023889 &    4014 & 4698 & 7 & (\ref{eqn:mdl:4})  & $   0.70{\pm}0.01$ & $   0.13{\pm}0.01$ & $  54.40{\pm}0.27$ & $      1.12{\pm}0.02$ & $   5.34{\pm}0.01$ & $   1.08{\pm}0.04$ & $   5.37{\pm}0.02$ &       --- \\
   5023931 &    7009 & 4718 & 8 & (\ref{eqn:mdl:4})  & $   0.64{\pm}0.02$ & $   0.19{\pm}0.02$ & $  51.14{\pm}0.62$ & $      1.01{\pm}0.06$ & $   4.91{\pm}0.04$ & $   1.18{\pm}0.06$ & $   4.83{\pm}0.03$ & $   5.10$ \\
   5023953 &    3011 & 4802 & 5 & (\ref{eqn:mdl:4})  & $   0.70{\pm}0.02$ & $   0.11{\pm}0.02$ & $  48.69{\pm}0.32$ & $      0.87{\pm}0.00$ & $   4.79{\pm}0.00$ & $   1.06{\pm}0.10$ & $   4.71{\pm}0.05$ & $   4.72$ \\
   5024043 &    8013 & 4796 & 5 & (\ref{eqn:mdl:4})  & $   0.74{\pm}0.01$ & $   0.10{\pm}0.01$ & $  56.24{\pm}0.24$ & $      1.09{\pm}0.00$ & $   5.60{\pm}0.00$ & $   1.05{\pm}0.05$ & $   5.62{\pm}0.03$ &       --- \\
   5024143 &    7005 & 4836 & 5 & (\ref{eqn:mdl:2})  & $   1.22{\pm}0.04$ & $   0.15{\pm}0.02$ & $ 119.76{\pm}1.13$ & $      1.21{\pm}0.05$ & $   9.65{\pm}0.05$ & $   1.22{\pm}0.03$ & $   9.64{\pm}0.03$ & $   9.73$ \\
   5024240 &    8007 & 4950 & 4 & (\ref{eqn:mdl:22}) & $   1.44{\pm}0.03$ & $   0.15{\pm}0.02$ & $ 150.29{\pm}2.19$ & $      1.26{\pm}0.06$ & $  11.97{\pm}0.06$ & $   1.26{\pm}0.03$ & $  11.97{\pm}0.03$ & $  11.99$ \\
   5024272 &    3003 & 4664 & 4 & (\ref{eqn:mdl:4})  & $   0.29{\pm}0.04$ & $   0.08{\pm}0.02$ & $  17.49{\pm}0.29$ & $      0.66{\pm}0.20$ & $   2.01{\pm}0.05$ & $   0.75{\pm}0.13$ & $   1.99{\pm}0.03$ & $   1.96$ \\
   5024297 &    8003 & 4673 & 5 & (\ref{eqn:mdl:4})  & $   0.58{\pm}0.01$ & $   0.08{\pm}0.01$ & $  46.29{\pm}0.22$ & $      1.02{\pm}0.06$ & $   4.58{\pm}0.03$ & $   1.03{\pm}0.04$ & $   4.57{\pm}0.02$ & $   4.73$ \\
   5024312 &   13002 & 4816 & 6 & (\ref{eqn:mdl:22}) & $   0.98{\pm}0.02$ & $   0.13{\pm}0.02$ & $  94.80{\pm}0.63$ & $      1.21{\pm}0.07$ & $   8.06{\pm}0.05$ & $   1.29{\pm}0.04$ & $   8.01{\pm}0.03$ & $   8.23$ \\
   5024327 &   11002 & 4790 & 6 & (\ref{eqn:mdl:4})  & $   0.71{\pm}0.02$ & $   0.18{\pm}0.02$ & $  45.02{\pm}0.36$ & $      0.92{\pm}0.07$ & $   4.75{\pm}0.04$ & $   0.97{\pm}0.05$ & $   4.73{\pm}0.03$ & $   4.70$ \\
   5024404 &    3004 & 4707 & 5 & (\ref{eqn:mdl:24}) & $   0.67{\pm}0.03$ & $   0.19{\pm}0.04$ & $  47.00{\pm}0.30$ & $      1.02{\pm}0.24$ & $   4.77{\pm}0.14$ & $   1.00{\pm}0.08$ & $   4.78{\pm}0.04$ & $   4.79$ \\
   5024405 &    4001 & 4775 & 5 & (\ref{eqn:mdl:22}) & $   1.03{\pm}0.02$ & $   0.19{\pm}0.03$ & $  95.36{\pm}0.89$ & $      1.17{\pm}0.05$ & $   8.26{\pm}0.04$ & $   1.18{\pm}0.02$ & $   8.25{\pm}0.01$ & $   8.28$ \\
   5024414 &    6002 & 5049 & 6 & (\ref{eqn:mdl:22}) & $   0.73{\pm}0.03$ & $   0.24{\pm}0.03$ & $  78.82{\pm}0.40$ & $      0.88{\pm}0.13$ & $   6.56{\pm}0.08$ & $   1.13{\pm}0.08$ & $   6.43{\pm}0.05$ & $   6.55$ \\
   5024476 &    1006 & 4945 & 0 & (\ref{eqn:mdl:22}) &                --- &                --- & $  66.80{\pm}0.54$ &                   --- &                --- &                --- & $   5.78{\pm}0.14$ & $   5.67$ \\
   5024512 &    3001 & 4826 & 7 & (\ref{eqn:mdl:22}) & $   0.82{\pm}0.01$ & $   0.09{\pm}0.01$ & $  73.56{\pm}0.39$ & $      1.09{\pm}0.06$ & $   6.69{\pm}0.04$ & $   1.14{\pm}0.06$ & $   6.65{\pm}0.04$ & $   6.87$ \\
   5024582 &    9002 & 4859 & 0 & (\ref{eqn:mdl:22}) &                --- &                --- & $  46.60{\pm}0.35$ &                   --- &                --- &                --- & $   4.74{\pm}0.12$ & $   4.79$ \\
   5024583 &    7003 & 4627 & 6 & (\ref{eqn:mdl:4})  & $   0.52{\pm}0.01$ & $   0.08{\pm}0.01$ & $  37.59{\pm}0.21$ & $      0.97{\pm}0.10$ & $   3.89{\pm}0.05$ & $   0.97{\pm}0.04$ & $   3.89{\pm}0.02$ & $   4.04$ \\
   5024601 &    4002 & 4716 & 7 & (\ref{eqn:mdl:24}) & $   0.55{\pm}0.02$ & $   0.13{\pm}0.02$ & $  33.12{\pm}0.43$ & $      0.85{\pm}0.02$ & $   3.71{\pm}0.01$ & $   0.93{\pm}0.06$ & $   3.68{\pm}0.03$ & $   3.67$ \\
   5024750 &    1004 & 4468 & 0 & (\ref{eqn:mdl:4})  &                --- &                --- & $  13.10{\pm}0.30$ &                   --- &                --- &                --- & $   1.78{\pm}0.05$ & $   1.81$ \\
   5024851 &    2008 & 4087 & 0 & (\ref{eqn:mdl:4})  &                --- &                --- & $   4.18{\pm}0.06$ &                   --- &                --- &                --- &                --- & $   0.75$ \\
   5024967 &    6009 & 4754 & 6 & (\ref{eqn:mdl:24}) & $   0.68{\pm}0.04$ & $   0.15{\pm}0.02$ & $  45.66{\pm}0.45$ & $      0.86{\pm}0.03$ & $   4.77{\pm}0.02$ & $   1.14{\pm}0.08$ & $   4.63{\pm}0.05$ & $   4.71$ \\
   5111718 &    8018 & 4932 & 6 & (\ref{eqn:mdl:2})  & $   1.29{\pm}0.02$ & $   0.23{\pm}0.03$ & $ 134.26{\pm}0.62$ & $      1.27{\pm}0.09$ & $  10.52{\pm}0.09$ & $   1.30{\pm}0.03$ & $  10.50{\pm}0.03$ & $  10.75$ \\
   5111940 &    5012 & 4741 & 7 & (\ref{eqn:mdl:22}) & $   0.66{\pm}0.01$ & $   0.12{\pm}0.01$ & $  53.12{\pm}0.33$ & $      1.04{\pm}0.05$ & $   5.18{\pm}0.03$ & $   1.14{\pm}0.06$ & $   5.14{\pm}0.03$ & $   5.32$ \\
   5111949 &    4011 & 4811 & 7 & (\ref{eqn:mdl:24}) & $   0.69{\pm}0.03$ & $   0.16{\pm}0.02$ & $  46.81{\pm}0.29$ & $      0.83{\pm}0.13$ & $   4.89{\pm}0.07$ & $   1.05{\pm}0.07$ & $   4.78{\pm}0.04$ & $   4.83$ \\
   5112072 &    9010 & 4929 & 5 & (\ref{eqn:mdl:24}) & $   1.25{\pm}0.02$ & $   0.13{\pm}0.02$ & $ 125.51{\pm}0.54$ & $      1.22{\pm}0.04$ & $  10.04{\pm}0.03$ & $   1.25{\pm}0.03$ & $  10.02{\pm}0.03$ & $  10.27$ \\
   5112288 &    2007 & 4793 & 7 & (\ref{eqn:mdl:4})  & $   0.66{\pm}0.02$ & $   0.14{\pm}0.02$ & $  47.85{\pm}0.36$ & $      0.88{\pm}0.18$ & $   4.82{\pm}0.10$ & $   1.14{\pm}0.07$ & $   4.70{\pm}0.04$ & $   4.86$ \\
   5112361 &    4008 & 4898 & 6 & (\ref{eqn:mdl:22}) & $   0.78{\pm}0.02$ & $   0.14{\pm}0.02$ & $  69.34{\pm}0.32$ & $      1.08{\pm}0.08$ & $   6.18{\pm}0.04$ & $   1.14{\pm}0.04$ & $   6.15{\pm}0.02$ & $   6.34$ \\
   5112373 &    5005 & 4779 & 4 & (\ref{eqn:mdl:4})  & $   0.65{\pm}0.03$ & $   0.10{\pm}0.02$ & $  44.14{\pm}0.32$ & $      0.87{\pm}0.20$ & $   4.67{\pm}0.12$ & $   0.87{\pm}0.08$ & $   4.66{\pm}0.05$ & $   4.63$ \\
   5112387 &    3007 & 4778 & 6 & (\ref{eqn:mdl:24}) & $   0.63{\pm}0.02$ & $   0.16{\pm}0.03$ & $  44.91{\pm}0.28$ & $      0.93{\pm}0.06$ & $   4.72{\pm}0.03$ & $   0.95{\pm}0.04$ & $   4.71{\pm}0.02$ & $   4.70$ \\
   5112401 &    3009 & 4732 & 0 & (\ref{eqn:mdl:24}) &                --- &                --- & $  36.45{\pm}0.33$ &                   --- &                --- &                --- & $   4.04{\pm}0.11$ & $   3.97$ \\
   5112467 &    6003 & 4778 & 0 & (\ref{eqn:mdl:22}) &                --- &                --- & $  45.96{\pm}0.27$ &                   --- &                --- &                --- & $   4.70{\pm}0.12$ & $   4.72$ \\
   5112481 &    1007 & 4126 & 0 & (\ref{eqn:mdl:24}) &                --- &                --- & $   5.59{\pm}0.08$ &                   --- &                --- &                --- &                --- & $   0.92$ \\
   5112491 &   10002 & 4802 & 5 & (\ref{eqn:mdl:24}) & $   0.72{\pm}0.02$ & $   0.13{\pm}0.02$ & $  44.58{\pm}0.29$ & $      0.92{\pm}0.19$ & $   4.71{\pm}0.09$ & $   0.99{\pm}0.04$ & $   4.68{\pm}0.02$ & $   4.62$ \\
   5112730 &    4005 & 4760 & 6 & (\ref{eqn:mdl:4})  & $   0.66{\pm}0.02$ & $   0.14{\pm}0.02$ & $  43.85{\pm}0.35$ & $      0.88{\pm}0.18$ & $   4.60{\pm}0.10$ & $   1.09{\pm}0.08$ & $   4.51{\pm}0.05$ & $   4.54$ \\
   5112734 &   12002 & 4607 & 6 & (\ref{eqn:mdl:22}) & $   0.53{\pm}0.01$ & $   0.14{\pm}0.02$ & $  40.49{\pm}0.24$ & $      1.02{\pm}0.11$ & $   4.13{\pm}0.06$ & $   1.00{\pm}0.05$ & $   4.14{\pm}0.03$ & $   4.23$ \\
   5112744 &    5011 & 4634 & 5 & (\ref{eqn:mdl:24}) & $   0.56{\pm}0.01$ & $   0.11{\pm}0.02$ & $  43.90{\pm}0.27$ & $      0.98{\pm}0.02$ & $   4.43{\pm}0.01$ & $   1.04{\pm}0.06$ & $   4.41{\pm}0.03$ & $   4.54$ \\
   5112786 &    5003 & 4245 & 3 & (\ref{eqn:mdl:24}) & $   0.18{\pm}0.01$ & $   0.12{\pm}0.01$ & $   8.01{\pm}0.10$ & $      0.01{\pm}0.19$ & $   1.10{\pm}0.04$ & $   0.01{\pm}0.19$ & $   1.10{\pm}0.04$ & $   1.22$ \\
   5112880 &    2004 & 4545 & 4 & (\ref{eqn:mdl:4})  & $   0.34{\pm}0.02$ & $   0.16{\pm}0.03$ & $  24.84{\pm}0.18$ & $      0.81{\pm}0.06$ & $   2.83{\pm}0.02$ & $   1.06{\pm}0.12$ & $   2.75{\pm}0.05$ & $   2.84$ \\
   5112938 &    2006 & 4773 & 6 & (\ref{eqn:mdl:24}) & $   0.69{\pm}0.03$ & $   0.12{\pm}0.02$ & $  44.84{\pm}0.29$ & $      1.06{\pm}0.21$ & $   4.69{\pm}0.12$ & $   1.13{\pm}0.07$ & $   4.66{\pm}0.04$ & $   4.66$ \\
   5112948 &    5007 & 4642 & 6 & (\ref{eqn:mdl:4})  & $   0.56{\pm}0.01$ & $   0.09{\pm}0.01$ & $  42.75{\pm}0.29$ & $      0.99{\pm}0.05$ & $   4.30{\pm}0.02$ & $   1.10{\pm}0.04$ & $   4.25{\pm}0.02$ & $   4.43$ \\
   5112950 &    3005 & 4714 & 5 & (\ref{eqn:mdl:24}) & $   0.62{\pm}0.02$ & $   0.11{\pm}0.02$ & $  41.18{\pm}0.34$ & $      0.88{\pm}0.19$ & $   4.38{\pm}0.09$ & $   1.08{\pm}0.08$ & $   4.30{\pm}0.04$ & $   4.33$ \\
   5112974 &    4009 & 4698 & 6 & (\ref{eqn:mdl:4})  & $   0.66{\pm}0.02$ & $   0.14{\pm}0.02$ & $  40.26{\pm}0.32$ & $      0.95{\pm}0.14$ & $   4.32{\pm}0.09$ & $   0.97{\pm}0.07$ & $   4.32{\pm}0.04$ & $   4.17$ \\
   5113041 &    4007 & 4607 & 6 & (\ref{eqn:mdl:4})  & $   0.52{\pm}0.01$ & $   0.11{\pm}0.02$ & $  37.82{\pm}0.26$ & $      1.02{\pm}0.01$ & $   3.97{\pm}0.00$ & $   0.97{\pm}0.07$ & $   3.99{\pm}0.03$ & $   4.07$ \\
   5113061 &    1014 & 4156 & 0 & (\ref{eqn:mdl:22}) &                --- &                --- & $   4.54{\pm}0.19$ &                   --- &                --- &                --- & $   0.83{\pm}0.02$ & $   0.80$ \\
   5113441 &   12016 & 4884 & 5 & (\ref{eqn:mdl:2})  & $   1.41{\pm}0.02$ & $   0.14{\pm}0.02$ & $ 155.35{\pm}0.91$ & $      1.31{\pm}0.00$ & $  11.68{\pm}0.00$ & $   1.28{\pm}0.02$ & $  11.70{\pm}0.02$ & $  11.91$ \\
   5200152 &    3021 & 4876 & 8 & (\ref{eqn:mdl:24}) & $   0.70{\pm}0.02$ & $   0.14{\pm}0.01$ & $  45.33{\pm}0.32$ & $      0.93{\pm}0.33$ & $   4.76{\pm}0.19$ & $   1.11{\pm}0.05$ & $   4.68{\pm}0.03$ & $   4.74$ \\
	\bottomrule
	\end{tabular}
	\begin{tablenotes}
		\scriptsize
		\item[1] ID from \citet{Milliman2014}.
		\item[2] From $(V-K_S)$ using \citet{CV2014} and a nominal reddening of $E(B-V)=0.15$.
		\item[3] Number of individual radial modes measured. If 0 only $\Dnu_\mathrm{global}$ was determined.
		\item[4] Derived using only the three central radial modes. See \citet{Kallinger2012}.
		\item[5] Empirically derived on a star-by-star basis using the distance modulus derived from eclipsing binaries.
	\end{tablenotes}
\end{threeparttable}
\end{table}

\clearpage
\begin{table}
\centering
\begin{threeparttable}
	\caption{Derived stellar properties.}
	\label{tab:table2}
	\tiny
	\begin{tabular}{rrrrrrrrrrrr}
	\toprule
	\mc{KIC} & \mc{WOCS} & \mc{$\Teff$} & \mc{Radius\tnote{1}}      & \mc{Mass\tnote{1}}  & \mc{$(m-M)_{V}$\tnote{2}} & \mc{$(m-M)_{V}$\tnote{3}} & \mc{CLASS}& \mc{PM\tnote{4}} & \mc{PRV\tnote{4}} & \mc{Mem\tnote{4}} & \mc{Mem} \\
	         &           & \mc{(K)}              & \mc{(\Rsun)} & \mc{(\Msun)} & \mc{(1)}         & \mc{(2)}         &           & \mc{(\%)} &  \mc{(\%)}  & \mc{Lit.} & \mc{Our} \\
	\midrule
   4937011 &    7017  & 4669 & $ 9.27{\pm}0.50$ & $0.71{\pm}0.08$ & $12.37$ & $12.31$ &   RC  &  0  & 94   &  SN  & SUM \\
   4937056 &    2012  & 4788 & $11.16{\pm}0.60$ & $1.74{\pm}0.20$ & $12.47$ & $12.41$ &   RC  & 98  & 94   &  BM  &  BM \\
   4937257 &    9015  & 4601 & $10.41{\pm}0.17$ & $1.08{\pm}0.04$ & $13.16$ & $13.09$ &   RC  & 81  & 84   &  SM  &  SN \\
   4937576 &    5016  & 4542 & $13.28{\pm}0.26$ & $1.67{\pm}0.08$ & $12.47$ & $12.40$ &  RGB  & 97  & 90   &  SM  &  SM \\
   4937770 &    9024  & 4930 & $ 7.09{\pm}0.35$ & $1.31{\pm}0.14$ & $12.04$ & $11.98$ &  RGB  &  5  & 93   &  SM  & SOM \\
   4937775 &    9026  & 5060 & $ 9.49{\pm}0.47$ & $2.51{\pm}0.25$ & $12.79$ & $12.74$ &  RGB  & 16  & 89   &  BM  & BOM \\
   5023732 &    5014  & 4588 & $14.29{\pm}0.35$ & $1.60{\pm}0.09$ & $12.45$ & $12.39$ &  RGB  & 99  & 93   &  SM  &  SM \\
   5023845 &    8010  & 4845 & $ 7.07{\pm}0.09$ & $1.61{\pm}0.05$ & $12.41$ & $12.35$ &  RGB  & 99  & 94   &  SM  &  SM \\
   5023889 &    4014  & 4698 & $ 9.39{\pm}0.12$ & $1.40{\pm}0.05$ & $11.20$ & $11.14$ &  RGB  &  0  & 93   &  SN  &  SN \\
   5023931 &    7009  & 4718 & $11.09{\pm}0.22$ & $1.84{\pm}0.09$ & $12.55$ & $12.49$ &  RGB  & 99  & 58   &  BM  &  BM \\
   5023953 &    3011  & 4802 & $11.28{\pm}0.27$ & $1.83{\pm}0.10$ & $12.33$ & $12.27$ &   RC  & 99  & 93   &  BM  & BOM \\
   5024043 &    8013  & 4796 & $ 8.93{\pm}0.12$ & $1.32{\pm}0.04$ & $12.01$ & $11.95$ &  RGB  &  0  & 94   &  SN  &  SN \\
   5024143 &    7005  & 4836 & $ 6.60{\pm}0.09$ & $1.55{\pm}0.06$ & $12.35$ & $12.29$ &  RGB  &  0  & 93   &  SN  &  SM \\
   5024240 &    8007  & 4950 & $ 5.44{\pm}0.09$ & $1.33{\pm}0.07$ & $12.32$ & $12.26$ &  RGB  & 94  & 88   &  BM  &  BM \\
   5024272 &    3003  & 4664 & $23.46{\pm}0.92$ & $2.80{\pm}0.25$ & $12.37$ & $12.31$ &  RGB  & 99  & 94   &  SM  & SOM \\
   5024297 &    8003  & 4673 & $11.14{\pm}0.14$ & $1.67{\pm}0.06$ & $12.46$ & $12.40$ &  RGB  & 99  & 85   &  SM  &  SM \\
   5024312 &   13002  & 4816 & $ 7.55{\pm}0.10$ & $1.60{\pm}0.05$ & $12.42$ & $12.37$ &  RGB  & 86  & 88   &  SM  &  SM \\
   5024327 &   11002  & 4790 & $10.80{\pm}0.17$ & $1.55{\pm}0.06$ & $12.39$ & $12.33$ &   RC  & 99  & 93   &  SM  &  SM \\
   5024404 &    3004  & 4707 & $10.93{\pm}0.23$ & $1.64{\pm}0.08$ & $12.43$ & $12.37$ &   RC  & 99  & 90   &  SM  &  SM \\
   5024405 &    4001  & 4775 & $ 7.13{\pm}0.09$ & $1.43{\pm}0.05$ & $12.33$ & $12.27$ &  RGB  & 98  & 91   &  SM  &  SM \\
   5024414 &    6002  & 5049 & $10.49{\pm}0.18$ & $2.63{\pm}0.11$ & $12.50$ & $11.93$ &   RC  & 99  & 94   &  BM  & BOM \\
   5024476 &    1006  & 4945 & $10.91{\pm}0.54$ & $2.38{\pm}0.24$ & $12.34$ & $11.90$ &   RC  & 99  & 94   &  BM  & BOM \\
   5024512 &    3001  & 4826 & $ 8.51{\pm}0.13$ & $1.57{\pm}0.06$ & $12.45$ & $12.39$ &  RGB  & 99  & 90   &  SM  &  SM \\
   5024582 &    9002  & 4859 & $11.21{\pm}0.56$ & $1.74{\pm}0.18$ & $12.46$ & $12.41$ &   RC  & 99  & 93   &  BM  &  BM \\
   5024583 &    7003  & 4627 & $12.43{\pm}0.16$ & $1.68{\pm}0.06$ & $12.47$ & $12.41$ &  RGB  & 95  & 93   &  SM  &  SM \\
   5024601 &    4002  & 4716 & $13.04{\pm}0.29$ & $1.65{\pm}0.09$ & $12.42$ & $12.36$ &   RC  & 76  & 91   &  SM  &  SM \\
   5024750 &    1004  & 4468 & $20.30{\pm}1.15$ & $1.54{\pm}0.19$ & $12.39$ & $12.32$ &  RGB  & 99  & 91   &  SM  &  SM \\
   5024851 &    2008  & 4087 & $36.87{\pm}0.64$ & $1.55{\pm}0.14$ & $12.42$ & $12.34$ &  RGB  & 99  & 93   &  BM  &  BM \\
   5024967 &    6009  & 4754 & $11.36{\pm}0.26$ & $1.73{\pm}0.09$ & $12.49$ & $12.43$ &   RC  & 98  & 89   &  SM  &  SM \\
   5111718 &    8018  & 4932 & $ 6.31{\pm}0.07$ & $1.60{\pm}0.05$ & $12.42$ & $12.36$ &  RGB  & 99  & 93   &  SM  &  SM \\
   5111940 &    5012  & 4741 & $10.21{\pm}0.16$ & $1.62{\pm}0.06$ & $12.46$ & $12.40$ &  RGB  & 98  & 94   &  SM  &  SM \\
   5111949 &    4011  & 4811 & $11.00{\pm}0.21$ & $1.67{\pm}0.07$ & $12.46$ & $12.40$ &   RC  & 99  & 92   &  SM  &  SM \\
   5112072 &    9010  & 4929 & $ 6.47{\pm}0.07$ & $1.57{\pm}0.05$ & $12.42$ & $12.36$ &  RGB  & 99  & 94   &  SM  &  SM \\
   5112288 &    2007  & 4793 & $11.62{\pm}0.22$ & $1.91{\pm}0.09$ & $12.56$ & $12.50$ &   RC  & 99  & 92   &  SM  &  SM \\
   5112361 &    4008  & 4898 & $ 9.47{\pm}0.11$ & $1.85{\pm}0.06$ & $12.45$ & $12.39$ &  RGB  & 98  & 88   &  BM  & BOM \\
   5112373 &    5005  & 4779 & $10.88{\pm}0.26$ & $1.54{\pm}0.08$ & $12.39$ & $12.33$ &   RC  & 98  & 93   &  SM  &  SM \\
   5112387 &    3007  & 4778 & $10.83{\pm}0.14$ & $1.55{\pm}0.05$ & $12.41$ & $12.35$ &   RC  & 99  & 94   &  SM  &  SM \\
   5112401 &    3009  & 4732 & $11.91{\pm}0.65$ & $1.51{\pm}0.17$ & $12.34$ & $12.28$ &   RC  & 99  & 94   &  SM  &  SM \\
   5112467 &    6003  & 4778 & $11.16{\pm}0.58$ & $1.68{\pm}0.18$ & $12.44$ & $12.39$ &   RC  & 99  & 94   &  SM  &  SM \\
   5112481 &    1007  & 4126 & $32.78{\pm}0.57$ & $1.64{\pm}0.15$ & $12.42$ & $12.34$ &  RGB  &  0  & 91   &  SN  &  SM \\
   5112491 &   10002  & 4802 & $10.92{\pm}0.16$ & $1.57{\pm}0.06$ & $12.36$ & $12.31$ &   RC  & 97  & 94   &  SM  &  SM \\
   5112730 &    4005  & 4760 & $11.54{\pm}0.27$ & $1.71{\pm}0.09$ & $12.45$ & $12.39$ &   RC  & 99  & 91   &  SM  &  SM \\
   5112734 &   12002  & 4607 & $11.83{\pm}0.19$ & $1.64{\pm}0.06$ & $12.41$ & $12.35$ &  RGB  &  0  & 87   &  SN  &  SM \\
   5112744 &    5011  & 4634 & $11.33{\pm}0.21$ & $1.63{\pm}0.07$ & $12.44$ & $12.37$ &  RGB  & 96  & 94   &  SM  &  SM \\
   5112786 &    5003  & 4245 & $31.56{\pm}2.10$ & $2.21{\pm}0.30$ & $12.73$ & $12.66$ &  RGB  & 99  & 86   &  SM  &  SM \\
   5112880 &    2004  & 4545 & $16.34{\pm}0.57$ & $1.90{\pm}0.14$ & $12.45$ & $12.38$ &  RGB  & 99  & 73   &  SM  & SOM \\
   5112938 &    2006  & 4773 & $11.06{\pm}0.22$ & $1.61{\pm}0.07$ & $12.42$ & $12.36$ &   RC  & 98  & 85   &  SM  &  SM \\
   5112948 &    5007  & 4642 & $11.86{\pm}0.18$ & $1.75{\pm}0.07$ & $12.49$ & $12.43$ &  RGB  & 98  & 92   &  SM  &  SM \\
   5112950 &    3005  & 4714 & $11.85{\pm}0.27$ & $1.69{\pm}0.09$ & $12.45$ & $12.39$ &   RC  & 99  & 94   &  SM  &  SM \\
   5112974 &    4009  & 4698 & $11.48{\pm}0.27$ & $1.55{\pm}0.08$ & $12.27$ & $12.21$ &   RC  & 94  & 93   &  SM  &  SM \\
   5113041 &    4007  & 4607 & $11.86{\pm}0.23$ & $1.54{\pm}0.07$ & $12.40$ & $12.33$ &  RGB  & 98  & 93   &  SM  &  SM \\
   5113061 &    1014  & 4156 & $31.49{\pm}2.04$ & $1.23{\pm}0.20$ & $12.19$ & $12.11$ &  RGB  & 99  & 94   &  SM  &  SM \\
   5113441 &   12016  & 4884 & $ 5.85{\pm}0.06$ & $1.58{\pm}0.05$ & $12.39$ & $12.33$ &  RGB  & 98  & 88   &  SM  &  SM \\
   5200152 &    3021  & 4876 & $11.20{\pm}0.19$ & $1.69{\pm}0.07$ & $12.48$ & $12.42$ &   RC  & 21  & 92   &  SM  &  SM \\
	\bottomrule
	\end{tabular}
	\begin{tablenotes}
		\scriptsize
		\item[1] Derived using a correction to \Dnu of 2.54\% for RGB stars and no correction to RC stars. For over- and undermassive stars and non-members a correction based on \fref{fig:corr} was employed.
		\item[2] Using bolometric corrections from \citet{CV2014}.
		\item[3] Using the calibration by \citet{DiBenedetto2005}.
		\item[4] From \citet{Milliman2014}.
	\end{tablenotes}
\end{threeparttable}
\end{table}

\clearpage
\begin{table}
\centering
\begin{threeparttable}
	\caption{All \Teff are in units of K.}
	\label{tab:table3}
	\tiny
	\begin{tabular}{rrrrrrrrrrrrr}
	\toprule
	\mc{KIC} & \mc{WOCS} & \mc{$V$\tnote{1}}  & \mc{$(V-K_s)$}   & \mc{$(b-y)$\tnote{2}}  & \mc{\Teff\tnote{3}} & \mc{\Teff\tnote{4}}  & \multicolumn{2}{c}{SAGA\tnote{2}} & \multicolumn{2}{c}{Lee-Brown\tnote{5}} & \multicolumn{2}{c}{APOGEE\tnote{6}} \\
	         &           &               &                  &               & $(V-K_s)$ & $(b-y)$ & \mc{\Teff} & \mc{\FeH} & \mc{\Teff} & \mc{\FeH} & \mc{\Teff} & \mc{\FeH} \\
	\midrule
   4937011  &    7017 & $ 13.59$ & $ 2.92$ & $ 0.77$ & $ 4669$ & $ 4680$ & $ 4730$ & $-0.36$ & $ 4636$ & $-0.05$ & $ 4620$ & $-0.03$ \\
   4937056  &    2012 & $ 13.12$ & $ 2.79$ & $ 0.72$ & $ 4788$ & $ 4823$ & $ 4789$ & $-0.26$ & $ 4709$ & $-0.13$ & $ 4761$ & $ 0.02$ \\
   4937257  &    9015 & $ 14.23$ & $ 3.02$ & $ 0.79$ & $ 4601$ & $ 4637$ & $ 4594$ & $-0.32$ & $ 4583$ & $ 0.06$ & $ 4650$ & $ 0.05$ \\
   4937576  &    5016 & $ 13.11$ & $ 3.09$ & $ 0.80$ & $ 4542$ & $ 4581$ & $ 4545$ & $-0.09$ & $ 4569$ & $-0.17$ & $ 4573$ & $ 0.05$ \\
   4937770  &    9024 & $ 13.49$ & $ 2.64$ & $ 0.73$ & $ 4930$ & $ 4789$ & $ 4925$ & $-0.35$ & $ 4952$ & $-0.11$ & $ 4877$ & $-0.01$ \\
   4937775  &    9026 & $ 13.45$ & $ 2.51$ & $ 0.68$ & $ 5060$ & $ 4961$ & $ 5104$ & $-0.88$ & $ 5120$ & $-0.18$ &     --- &     --- \\
   5023732  &    5014 & $ 12.86$ & $ 3.02$ & $ 0.76$ & $ 4588$ & $ 4702$ & $ 4565$ & $ 0.40$ & $ 4539$ & $-0.05$ & $ 4541$ & $ 0.05$ \\
   5023845  &    8010 & $ 13.98$ & $ 2.73$ & $ 0.69$ & $ 4845$ & $ 4936$ & $ 4838$ & $ 0.09$ & $ 4779$ & $-0.14$ &     --- &     --- \\
   5023889  &    4014 & $ 12.36$ & $ 2.89$ & $ 0.74$ & $ 4698$ & $ 4773$ & $ 4646$ & $ 0.32$ & $ 4604$ & $-0.20$ & $ 4568$ & $ 0.15$ \\
   5023931  &    7009 & $ 13.31$ & $ 2.87$ & $ 0.73$ & $ 4718$ & $ 4793$ & $ 4707$ & $-0.04$ & $ 4646$ & $-0.57$ & $ 4648$ & $ 0.06$ \\
   5023953  &    3011 & $ 12.94$ & $ 2.77$ & $ 0.71$ & $ 4802$ & $ 4884$ & $ 4772$ & $ 0.12$ &     --- &     --- & $ 4780$ & $ 0.14$ \\
   5024043  &    8013 & $ 13.13$ & $ 2.78$ & $ 0.70$ & $ 4796$ & $ 4891$ & $ 4832$ & $-0.29$ & $ 4806$ & $-0.28$ & $ 4676$ & $-0.19$ \\
   5024143  &    7005 & $ 14.08$ & $ 2.74$ & $ 0.69$ & $ 4836$ & $ 4929$ & $ 4876$ & $-0.09$ & $ 4839$ & $-0.15$ & $ 4818$ & $ 0.06$ \\
   5024240  &    8007 & $ 14.32$ & $ 2.62$ & $ 0.68$ & $ 4950$ & $ 4990$ & $ 4943$ & $-0.40$ & $ 4950$ & $-0.05$ & $ 4864$ & $-0.09$ \\
   5024272  &    3003 & $ 11.58$ & $ 2.93$ & $ 0.78$ & $ 4664$ & $ 4633$ & $ 4568$ & $ 0.16$ & $ 4519$ & $-0.37$ &     --- &     --- \\
   5024297  &    8003 & $ 13.28$ & $ 2.92$ & $ 0.75$ & $ 4673$ & $ 4753$ & $ 4619$ & $ 0.11$ & $ 4623$ & $-0.15$ &     --- &     --- \\
   5024312  &   13002 & $ 13.89$ & $ 2.76$ & $ 0.70$ & $ 4816$ & $ 4919$ & $ 4795$ & $-0.02$ & $ 4781$ & $-0.08$ &     --- &     --- \\
   5024327  &   11002 & $ 13.11$ & $ 2.79$ & $ 0.71$ & $ 4790$ & $ 4860$ & $ 4849$ & $ 0.05$ & $ 4731$ & $-0.19$ &     --- &     --- \\
   5024404  &    3004 & $ 13.24$ & $ 2.88$ & $ 0.75$ & $ 4707$ & $ 4750$ & $ 4684$ & $-0.24$ & $ 4692$ & $-0.07$ &     --- &     --- \\
   5024405  &    4001 & $ 13.97$ & $ 2.81$ & $ 0.73$ & $ 4775$ & $ 4819$ & $ 4714$ & $-0.13$ & $ 4739$ & $-0.08$ &     --- &     --- \\
   5024414  &    6002 & $ 12.95$ & $ 2.52$ & $ 0.66$ & $ 5049$ & $ 5045$ & $ 4996$ & $ 0.01$ & $ 4946$ & $-0.09$ &     --- &     --- \\
   5024476  &    1006 & $ 12.83$ & $ 2.62$ & $ 0.68$ & $ 4945$ & $ 4979$ & $ 4950$ & $-0.26$ &     --- &     --- & $ 4880$ & $ 0.11$ \\
   5024512  &    3001 & $ 13.64$ & $ 2.75$ & $ 0.70$ & $ 4826$ & $ 4912$ & $ 4740$ & $-0.01$ & $ 4790$ & $-0.16$ &     --- &     --- \\
   5024582  &    9002 & $ 13.01$ & $ 2.71$ & $ 0.69$ & $ 4859$ & $ 4929$ & $ 4822$ & $-0.12$ &     --- &     --- &     --- &     --- \\
   5024583  &    7003 & $ 13.12$ & $ 2.98$ & $ 0.76$ & $ 4627$ & $ 4708$ & $ 4609$ & $ 0.13$ & $ 4567$ & $-0.06$ &     --- &     --- \\
   5024601  &    4002 & $ 12.83$ & $ 2.87$ & $ 0.74$ & $ 4716$ & $ 4776$ & $ 4678$ & $ 0.02$ & $ 4677$ & $-0.09$ &     --- &     --- \\
   5024750  &    1004 & $ 12.22$ & $ 3.18$ &     --- & $ 4468$ &     --- &     --- &     --- & $ 4393$ & $-0.20$ &     --- &     --- \\
   5024851  &    2008 & $ 11.69$ & $ 3.79$ & $ 0.97$ & $ 4087$ & $ 4108$ & $ 4063$ & $ 0.64$ &     --- &     --- & $ 4174$ & $ 0.09$ \\
   5024967  &    6009 & $ 13.15$ & $ 2.83$ & $ 0.72$ & $ 4754$ & $ 4823$ & $ 4749$ & $-0.15$ & $ 4744$ & $ 0.00$ & $ 4705$ & $ 0.07$ \\
   5111718  &    8018 & $ 14.12$ & $ 2.64$ & $ 0.67$ & $ 4932$ & $ 4997$ & $ 4914$ & $ 0.03$ & $ 4890$ & $ 0.00$ & $ 4860$ & $ 0.14$ \\
   5111940  &    5012 & $ 13.37$ & $ 2.84$ & $ 0.73$ & $ 4741$ & $ 4819$ & $ 4718$ & $ 0.11$ & $ 4688$ & $-0.09$ & $ 4632$ & $ 0.06$ \\
   5111949  &    4011 & $ 13.11$ & $ 2.76$ & $ 0.72$ & $ 4811$ & $ 4850$ & $ 4776$ & $ 0.12$ & $ 4731$ & $-0.03$ & $ 4670$ & $ 0.10$ \\
   5112072  &    9010 & $ 14.07$ & $ 2.64$ & $ 0.67$ & $ 4929$ & $ 5004$ & $ 4901$ & $ 0.03$ & $ 4865$ & $-0.20$ & $ 4841$ & $ 0.04$ \\
   5112288  &    2007 & $ 13.12$ & $ 2.78$ & $ 0.74$ & $ 4793$ & $ 4770$ & $ 4787$ & $-0.14$ & $ 4716$ & $-0.07$ &     --- &     --- \\
   5112361  &    4008 & $ 13.31$ & $ 2.67$ & $ 0.70$ & $ 4898$ & $ 4912$ & $ 4869$ & $-0.39$ & $ 4900$ & $-0.35$ &     --- &     --- \\
   5112373  &    5005 & $ 13.11$ & $ 2.80$ & $ 0.73$ & $ 4779$ & $ 4813$ & $ 4729$ & $ 0.00$ & $ 4681$ & $-0.02$ & $ 4664$ & $ 0.06$ \\
   5112387  &    3007 & $ 13.14$ & $ 2.80$ & $ 0.73$ & $ 4778$ & $ 4799$ & $ 4746$ & $-0.09$ & $ 4709$ & $-0.13$ & $ 4684$ & $ 0.05$ \\
   5112401  &    3009 & $ 12.93$ & $ 2.85$ & $ 0.73$ & $ 4732$ & $ 4819$ & $ 4720$ & $ 0.01$ & $ 4722$ & $-0.13$ &     --- &     --- \\
   5112467  &    6003 & $ 13.11$ & $ 2.80$ & $ 0.71$ & $ 4778$ & $ 4853$ & $ 4785$ & $ 0.05$ & $ 4714$ & $-0.11$ &     --- &     --- \\
   5112481  &    1007 & $ 11.86$ & $ 3.71$ &     --- & $ 4126$ &     --- &     --- &     --- & $ 4149$ & $-0.21$ & $ 4265$ & $ 0.05$ \\
   5112491  &   10002 & $ 13.04$ & $ 2.77$ & $ 0.71$ & $ 4802$ & $ 4857$ & $ 4782$ & $-0.17$ & $ 4752$ & $-0.06$ & $ 4755$ & $ 0.03$ \\
   5112730  &    4005 & $ 13.07$ & $ 2.82$ & $ 0.75$ & $ 4760$ & $ 4753$ & $ 4733$ & $-0.24$ & $ 4714$ & $-0.14$ & $ 4693$ & $ 0.09$ \\
   5112734  &   12002 & $ 13.20$ & $ 3.00$ & $ 0.75$ & $ 4607$ & $ 4727$ & $ 4683$ & $-0.01$ & $ 4639$ & $-0.15$ & $ 4585$ & $ 0.09$ \\
   5112744  &    5011 & $ 13.28$ & $ 2.97$ & $ 0.76$ & $ 4634$ & $ 4692$ & $ 4643$ & $-0.11$ & $ 4645$ & $-0.22$ & $ 4601$ & $ 0.03$ \\
   5112786  &    5003 & $ 12.01$ & $ 3.51$ & $ 0.89$ & $ 4245$ & $ 4315$ & $ 4249$ & $ 0.11$ & $ 4233$ & $-0.18$ &     --- &     --- \\
   5112880  &    2004 & $ 12.63$ & $ 3.08$ & $ 0.79$ & $ 4545$ & $ 4605$ & $ 4497$ & $ 0.10$ & $ 4500$ & $-0.04$ & $ 4568$ & $ 0.05$ \\
   5112938  &    2006 & $ 13.11$ & $ 2.81$ & $ 0.73$ & $ 4773$ & $ 4796$ & $ 4756$ & $-0.08$ & $ 4724$ & $-0.11$ &     --- &     --- \\
   5112948  &    5007 & $ 13.22$ & $ 2.96$ & $ 0.76$ & $ 4642$ & $ 4715$ & $ 4643$ & $ 0.02$ & $ 4646$ & $-0.13$ &     --- &     --- \\
   5112950  &    3005 & $ 13.08$ & $ 2.87$ & $ 0.75$ & $ 4714$ & $ 4750$ & $ 4756$ & $-0.01$ & $ 4666$ & $-0.16$ & $ 4667$ & $ 0.09$ \\
   5112974  &    4009 & $ 12.99$ & $ 2.89$ & $ 0.74$ & $ 4698$ & $ 4773$ & $ 4650$ & $-0.24$ & $ 4710$ & $-0.07$ & $ 4686$ & $ 0.04$ \\
   5113041  &    4007 & $ 13.18$ & $ 3.00$ & $ 0.75$ & $ 4607$ & $ 4724$ & $ 4587$ & $ 0.03$ & $ 4618$ & $-0.13$ & $ 4598$ & $ 0.01$ \\
   5113061  &    1014 & $ 11.65$ & $ 3.66$ & $ 0.96$ & $ 4156$ & $ 4141$ & $ 4161$ & $ 0.16$ & $ 4113$ & $-0.28$ &     --- &     --- \\
   5113441  &   12016 & $ 14.32$ & $ 2.69$ & $ 0.68$ & $ 4884$ & $ 4990$ & $ 4837$ & $ 0.20$ & $ 4882$ & $ 0.03$ & $ 4842$ & $ 0.13$ \\
   5200152  &    3021 & $ 13.00$ & $ 2.69$ & $ 0.72$ & $ 4876$ & $ 4846$ & $ 4849$ & $-0.31$ & $ 4845$ & $ 0.00$ & $ 4753$ & $ 0.08$ \\
	\bottomrule
	\end{tabular}
	\begin{tablenotes}
		\scriptsize
		\item[1] From \citet{Milliman2014} except for KIC~5112481 and KIC~5113061 which are from \citet{Hole2009}.
		\item[2] \citet{Casagrande2014}.
		\item[3] From $(V-K_S)$ using \citet{CV2014} and a nominal reddening of $E(B-V)=0.15$.
		\item[4] Using $(b-y)$ from \citet{Casagrande2014}, $E(B-V)=0.15$ and colour-\Teff calibration by \citet{Ramirez2005}.
		\item[5] \citet{Lee-Brown2015}.
		\item[6] \citet{Pinsonneault2014}.
    \end{tablenotes}
\end{threeparttable}
\end{table}

\twocolumn
\bibliography{library}
\end{document}